\documentclass[12pt]{article}
\usepackage{amsthm,amsmath,amssymb,bm,bbm,graphicx,natbib,color}
\usepackage{enumerate}
\usepackage{float}
\usepackage{subfigure}
\usepackage{url} % not crucial - just used below for the URL 
\usepackage[colorlinks=true, citecolor=blue]{hyperref}
\usepackage{multirow,booktabs}

\newcommand{\blind}{1}

\newcommand{\tE}{\mathbb{E}}

\DeclareMathOperator*{\argmin}{arg\,min}

\newcommand{\bX}{\mathbf{X}}
\newcommand{\bx}{\mathbf{x}}
\newcommand{\bY}{\mathbf{Y}}
\newcommand{\by}{\mathbf{y}}
\newcommand{\bZ}{\mathbf{Z}}
\newcommand{\bz}{\mathbf{z}}

\newcommand{\bc}{\mathbf{c}}

\numberwithin{equation}{section}
\theoremstyle{plain}
\newtheorem{theorem}{Theorem}
\newtheorem{lemma}{Lemma}

\newtheorem{proposition}{Proposition}
\newtheorem{algorithm}{Algorithm}

\begin{document}

\def\spacingset#1{\renewcommand{\baselinestretch}%
{#1}\small\normalsize} \spacingset{1}

%%%%%%%%%%%%%%%%%%%%%%%%%%%%%%%%%%%%%%%%%%%%%%%%%%%%%%%%%%%%%%%%%%%%%%%%%%%%%%

\if1\blind
{
  \title{\bf Density Regression with Conditional Support Points}
  \author{Yunlu Chen\thanks{Yunlu Chen is now a PhD student at Northwestern University.}\hspace{.2cm}\\
    School of Data Science, Fudan University\\
    and \\
    Nan Zhang\thanks{
    Corresponding author} \\
    School of Data Science, Fudan University}
  \date{}
  \maketitle
} \fi

\if0\blind
{
  \bigskip
  \bigskip
  \bigskip
  \begin{center}
    {\LARGE\bf Density Regression with Conditional Support Points}
\end{center}
  \medskip
} \fi

\bigskip
\begin{abstract}
Density regression characterizes the conditional density of the response variable given the covariates, and provides much more information than the commonly used conditional mean or quantile regression. However, it is often computationally prohibitive in applications with massive data sets, especially when there are multiple covariates. In this paper, we develop a new data reduction approach for the density regression problem using conditional support points. After obtaining the representative data, we exploit the penalized likelihood method as the downstream estimation strategy. Based on the connections among the continuous ranked probability score, the energy distance, the $L_2$ discrepancy and the symmetrized Kullback-Leibler distance, we investigate the distributional convergence of the representative points and establish the rate of convergence of the density regression estimator. The usefulness of the methodology is illustrated by modeling the conditional distribution of power output given multivariate environmental factors using a large scale wind turbine data set. Supplementary materials for this article are available online.
\end{abstract}

\noindent%
{\it Keywords:}  Density regression; Data reduction; Energy distance; Penalized likelihood estimation; Kullback-Leibler distance; $L_2$ discrepancy.
\vfill

\newpage
\spacingset{1.5} % DON'T change the spacing!

%%%%%%%%%%%%%%%%%%%%%%%%%%%%%%%%%%%%%%%%
\section{Introduction}\label{int}

Density regression, also known as conditional density estimation, is an appealing statistical method to describe the distributional change of a response variable $Y\in\mathcal Y$ according to covariates $\bX\in\mathcal X\subset \mathbb R^d$. Compared with conditional mean or quantile regression, density regression requires a less restrictive assumption on residual distribution. In particular, when the response distribution of interest varies with covariates it is inappropriate to assume that the residual distribution in regression models is constant over $\mathcal X$. Its ability to fully characterize the conditional distribution makes density regression attractive in applications. For example, in the wind energy industry, engineers use the power curve to assess the operational performance of wind turbines. The power curve describes the functional relationship between the power output generated by a turbine and the wind speed. Practical evidence has shown that changes in environmental factors, such as wind speed, wind direction, and air density, may lead to varying skewness and moving modes \citep{jeon2012using, lee2015power}.

With the proliferation of massive data, conventional statistical methods are often prohibitive in practice due to their high computational cost. In recent years, there has been a growing interest in the development of data reduction for statistical models. For example, subsampling can efficiently extract useful information from data, and efficient algorithms have been devised for linear regression \citep{drineas2011faster, ma2015statistical, wang2019information}, logistic regression \citep{wang2018optimal, cheng2020information} and generalized linear models \citep{ai2018optimal}. Recently, \cite{joseph2021supervised} developed a sequential procedure which integrates response information to supervise data compression. It does not rely on parametric modeling assumptions and is robust to various modeling choices. Another probabilistic data compression technique is sketching, which uses random projections to generate a smaller surrogate dataset \citep{mahoney2011randomized, woodruff2014sketching}. Most available results are under the context of linear regression with ordinary least squares while both estimation and inference problems have been investigated \citep{pilanci2016iterative, raskutti2016statistical, ahfock2021statistical}. In the context of density estimation, data reduction is understood as selecting representative points from the full data. Beyond simply drawing a random subsample from the full data, various efficient algorithms have been proposed to choose representative points by minimizing certain statistical potential measures, such as the energy design \citep{joseph2015sequential}, the Riesz energy \citep{borodachov2014low}, and the energy distance \citep{mak2018support}. 

Our focus in this paper is on density regression for large datasets. Let $\mathbb D_N=\{(\bx_i, y_i): i=1,\dots,N\}$ be independent and identically distributed observations from a joint density $f(\bx,y)$ on the product domain $\mathcal X \times\mathcal Y$. The primary goal is to estimate the conditional distribution function $F(y \mid  \bx )$ or equivalently the conditional density function $f(y \mid \bx)$ based on $\mathbb D_N$. There is a large body of literature on conditional density estimation. One popular approach is the kernel-based methods. \cite{rosenblatt1969conditional} and \cite{hyndman1996estimating} considered the conventional kernel estimator for conditional density estimation and investigated its asymptotic properties. \cite{fan1996estimation} and \cite{hyndman2002nonparametric} developed a direct approach using locally parametric regression. The second approach uses splines. \cite{kooperberg1991study} proposed the log-spline model to enforce positivity and unity in density estimation. \cite{stone1994use,stone1997polynomial} considered using tensor products of polynomial splines to obtain conditional log density estimates. \cite{gu1995smoothing} adopted the smoothing spline analysis of variance (ANOVA) models for multivariate cases. From the Bayesian perspective, mixtures of experts model \citep{jacobs1991adaptive,jordan1994hierarchical} inspired recent research advancement using flexible priors \citep{dunson2007bayesian, chung2009nonparametric}. In terms of time complexity, computing a kernel density estimate requires $O(N^2)$ kernel function evaluations, and the spline-based approaches take $O(N^3)$ flops to invert large matrices of size $N$, let alone computationally intensive bandwidth selection and cross-validation procedures. To sum up, the computational burden of classical nonparametric methods grows quickly with the full sample size $N$, and it is computationally prohibitive to directly apply those methods to large datasets. In this paper, we focus on smoothing spline approach for density regression \citep{gu1995smoothing,jeon2006effective} since it has shown potential in multivariate modeling with ANOVA structure.

We develop a data reduction approach to alleviate the computational burden of density regression. Extending the idea of support points \citep{mak2018support} which is originally designed for compacting a continuous probability distribution into representative data points, we first partition the multivariate covariate space and then select support points for the conditional distribution within each partition. After combining the selected representative points, we exploit the penalized likelihood estimation \citep{gu1995smoothing, gu2013nonparametric} as the downstream method to obtain the final conditional density estimator. The continuous ranked probability score (CRPS) \citep{matheson1976scoring, gneiting2007strictly} is used to compare the estimated cumulative distribution function $\widehat F$ with the observed value. We show that the expected value of CRPS is closely related to the so-called energy distance \citep{szekely2004testing, szekely2013energy} which is minimized by support points within each partition. Furthermore, we rigorously investigate the statistical properties of our method. In particular, we obtain the distributional convergence of the proposed estimator to the desired distribution. Moreover, we use the symmetrized Kullback-Leibler distance to bound the $L_2$ discrepancy, and establish the convergence rate of the estimation error.

\begin{figure}[h!]
\centering
\subfigure{
\begin{minipage}[b]{0.3\linewidth}
\includegraphics[width=1\linewidth]{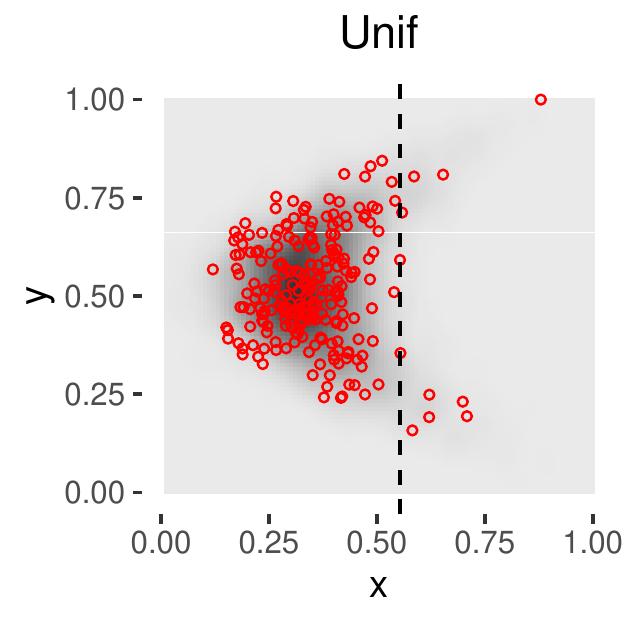}
\end{minipage}}
\subfigure{
\begin{minipage}[b]{0.3\linewidth}
\includegraphics[width=1\linewidth]{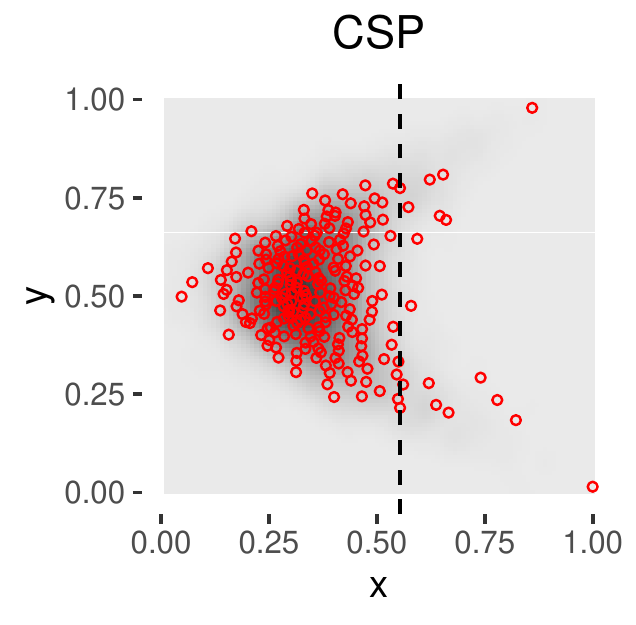}
\end{minipage}}
\subfigure{
\begin{minipage}[b]{0.3\linewidth}
\includegraphics[width=1\linewidth]{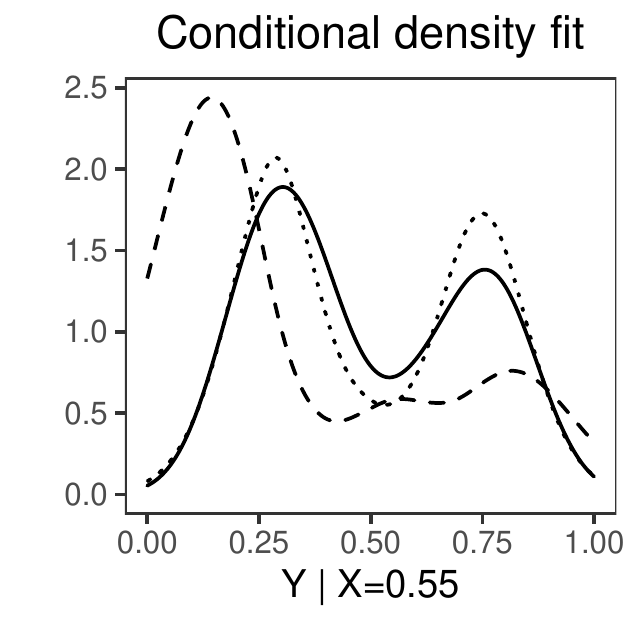}
\end{minipage}}
\caption{The banana distribution. Left and middle panels: shaded contour is the true joint density, red points are the representative points selected by the uniform subsampling and our method, dashed line indicates the location examined in the right panel. Right panel: the true conditional density (solid), density fits by the uniform subsampling (dashed), and by our method (dotted).}
\label{fig1}
\end{figure}
\begin{figure}[h!]
\centering
\subfigure{
\begin{minipage}[b]{0.3\linewidth}
\includegraphics[width=1\linewidth]{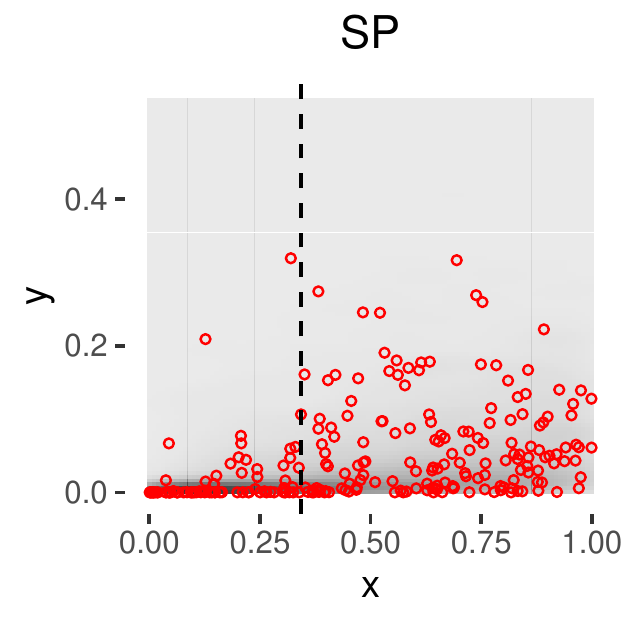}
\end{minipage}}
\subfigure{
\begin{minipage}[b]{0.3\linewidth}
\includegraphics[width=1\linewidth]{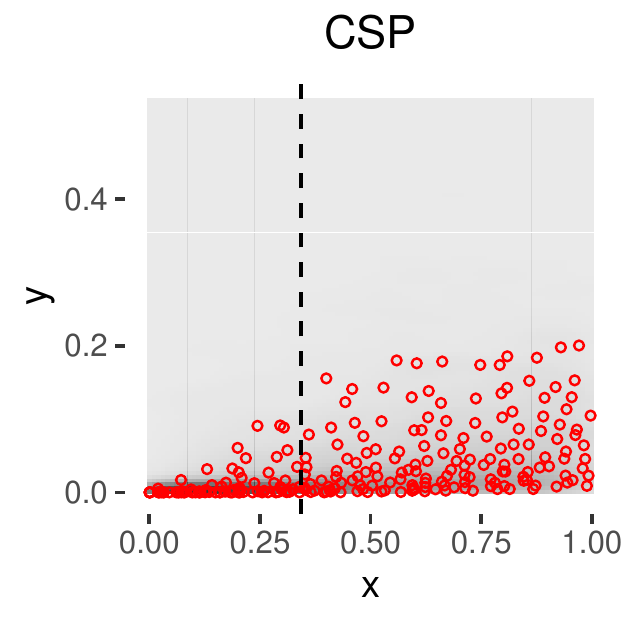}
\end{minipage}}
\subfigure{
\begin{minipage}[b]{0.3\linewidth}
\includegraphics[width=1\linewidth]{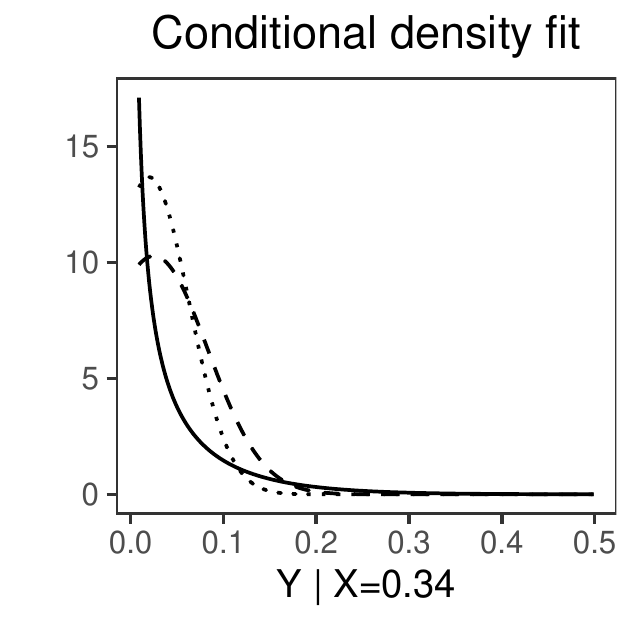}
\end{minipage}}
\caption{The conditional Beta distribution: $X$ is $\textrm{Uniform}(0,1)$ and  $Y\mid X$ is $\textrm{Beta}(X, X^2+10)$. Left and middle panels: shaded contour is the true joint density, red points are the representative points selected by the vanilla support points and our method, dashed line indicates the location examined in the right panel. Right panel: the true conditional density (solid), density fits by the vanilla support points (dashed), and by our method (dotted).
}
\label{fig2}
\end{figure}
The novelty of our method is that it respects the conditional structure of the problem and distinguishes the different roles played by the covariates and response variables in the context of density regression. Its optimality thus follows from that of the support points in the conditional sense. 
For illustration, we use two toy examples to make a comparison of our method with the uniform subsampling and the vanilla support points, respectively.
In Figure~\ref{fig1}, the left and middle panels plot the representative points selected by the uniform subsampling and our method. Both methods can represent the truth well but ours exhibits a clear space-filling property \citep{santner2003design}, that is, the points are concentrated in regions with high densities while they are well spread throughout the product domain. In the right panel of Figure~\ref{fig1}, the underlying true conditional density is bi-modal. It is recovered by our method while the uniform subsampling fails because few data are selected in this area.
Although the uniform subsampling is easy to implement, it is inefficient when samples are sparse. 
Figure~\ref{fig2} compares the vanilla support points with our method when the conditional distribution of $Y$ given $X$ is a Beta distribution. We provide an additional figure in the Supplementary Material by zooming into $Y\in[0, 0.1]$. When the covariate $X$ increases, the conditional distribution of $Y$ becomes more spread out. The left and middle panels show that the vanilla support points are scattered in the product domain $\mathcal X\times \mathcal Y$ while the points selected by our method follows the conditional structure more closely. In the right panel, the conditional density fit provided by our method is more accurate. In fact, the vanilla support points are designed with optimality for joint distribution. They are, however, not necessarily optimal for density regression because the conditional structure is ignored. 

The rest of the paper is organized as follows. Section \ref{methodology} presents the framework of data reduction and density regression and introduces the downstream penalized likelihood estimation. Section \ref{algorithm} proposes our algorithms of conditional support points for density regression. Section \ref{theoretical} proves the theoretical properties of our method. Section \ref{simulation} and Section \ref{real} demonstrate the practical effectiveness of conditional support points via simulation and application. Section~\ref{conclusion} concludes the article. For notational convenience, we sometimes omit the subscript of distribution or density function without causing confusion. 

%%%%%%%%%%%%%%%%%%%%%%%%%%%%%%%%%%%%%%%%
\section{Methodology}\label{methodology}

\subsection{Support points}
We first review the basics of support points \citep{mak2018support} which is the building block of our method. Support points are used to compact a continuous probability distribution into a set of representative points. They are obtained by optimizing a statistical potential measure called the energy distance \citep{szekely2004testing, szekely2013energy}. Let $F$ and $G$ be two cumulative distribution functions on a nonempty domain $\mathcal{Z} \subseteq \mathbb{R}^{p}$ with finite means. The energy distance between $F$ and $G$ is defined as
\begin{equation}\label{def:energy}
E(F, G) = 2 \mathbb{E} \|\bY-\bZ\|_{2}-\mathbb{E}\|\bY-\bY'\|_{2}-\mathbb{E}\|\bZ-\bZ'\|_{2},
\end{equation}
where $\bY, \bY'$ and $\bZ, \bZ'$ are independent and identically distributed copies from $F$ and $G$, respectively.
An important property of the energy distance is that $E(F, G)\geq 0$, and $E(F, G)=0$ if and only if $F=G$. Let $G=F_n$ be the empirical distribution function for $\{\bz_i\}_{i=1}^n\subset\mathcal Z$. 
Support points are then defined as
$$
\{\by_i^{*}\}_{i=1}^{n} = \underset{\bz_{1}, \dots, \bz_{n}}{\operatorname{argmin}}\, E\left(F, F_{n}\right)=\underset{\bz_{1}, \dots, \bz_{n}}{\operatorname{argmin}}\biggl\{\frac{2}{n} \sum_{j=1}^{n} \tE \|\bz_{j}-\bY\|_2- \frac{1}{n^{2}}
\sum_{i=1}^{n}\sum_{j=1}^{n}\|\bz_{i}-\bz_{j}\|_2\biggr\}.
$$
The two terms in the above display not only encourage support points to represent the truth distribution but also force them to be spread out within the entire domain. This is called the space-filling property in the design of experiments \citep{santner2003design}.
In practice, when $\{\by_m\}_{m=1}^N$ are independent and identically observed according to $F$, the above energy distance can be approximated by
$$
\widehat{E}\left(F, F_{n}\right)
=\frac{2}{nN} \sum_{j=1}^{n} \sum_{m=1}^{N} \|\bz_j-\by_m\|_2- \frac{1}{n^{2}}
\sum_{i=1}^{n}\sum_{j=1}^{n}\|\bz_i-\bz_j\|_2.
$$

Such an optimization problem can be formulated as a difference-of-convex program. Efficient algorithms for generating support points are available with theoretical guarantees based on the convex-concave procedures \citep{yuille2003concave}. The computational cost of support points is $O(Nnp)$ when the objective gap from the stationary solution is fixed, where $p$ is the dimensionality of $\mathcal Z$. It can be further reduced by parallel computation. See \cite{mak2018support} for details.

\subsection{Continuous ranked probability score}\label{sec:crps}

Density regression aims at estimating the conditional density function $f(y\mid\bx)$ or the conditional cumulative distribution function $F(y\mid \bx)$ based on independent and identically distributed observations $\mathbb D_N=\{(\bx_i, y_i): i=1,\dots,N\}$. The continuous ranked probability score (CRPS) \citep{matheson1976scoring} is a commonly used scoring rule for probabilistic density forecast \citep{gneiting2007strictly,gneiting2011comparing,ehm2016quantiles}. It compares the estimated cumulative distribution function with the observations via the binary events $\{1(y>y_i): i=1,\dots, N\}$. For a generic density regression estimator $\widehat F$, the average CRPS over $\mathbb D_N$ compares the estimated cumulative distribution function with the observed values, which is defined as
$$
\textrm{CRPS}(\widehat F)=N^{-1} \sum_{i=1}^N \int_{\mathcal{Y}}\left\{\widehat F(y\mid  \bx_i)-1(y>y_i)\right\}^2\,\textrm{d}y,
$$
where $\widehat F(y\mid  \bx_i)$ is the estimated cumulative distribution function given $\bx_i$.

As a statistical measure of density regression estimator, CRPS is closely related to the energy distance and the $L_2$ discrepancy. 
First, \cite{gneiting2007strictly} pointed out the connection between CRPS and the energy distance from the statistical energy perspective \citep{szekely2013energy}.
Second, the squared $L_2$ discrepancy between the univariate distribution functions $F$ and $G$ is 
\begin{equation}\label{eq:equiv}
D_2^2(F, G)=\int_{\mathcal Z} \{F(z)-G(z)\}^{2} \,\textrm{d}z.
\end{equation}
It can be shown that $D_2^2(F, G)$ equals one half of the energy distance $E(F, G)$, that is, using the notations in \eqref{def:energy} for a univariate case, we have
\begin{equation}\label{eq:d2energy}
D_2^2(F, G)= \mathbb{E} \|Y-Z\|_{2}-\frac{1}{2}\mathbb{E}\|Y-Y'\|_{2}-\frac{1}{2}\mathbb{E}\|Z-Z'\|_{2}.
\end{equation}
Under the case of the conditional distribution, the two connections still hold because $\widehat F$ and $F$ are both univariate. With slight abuses of notation, define the squared conditional $L_2$ discrepancy over $\mathbb D_N$ by
\begin{equation}\label{eq:d2}
D_2^2(\widehat F_{Y\mid \bX},F_{Y\mid \bX})=N^{-1}\sum_{i=1}^N\int_{\mathcal{Y}}\left\{ \widehat F_{Y\mid \bX}(y\mid  \bx _i)-F_{Y\mid \bX}(y\mid  \bx _i)\right\}^2\,\textrm{d}y.
\end{equation}
We show in the following that the expected CRPS of the density regression estimator consists of two parts. One is the expected CRPS of the truth while the other part is the expected squared conditional $L_2$ discrepancy between the truth and the estimator. Later we develop data reduction procedures based on this decomposition.
\begin{proposition}\label{prop:crps}
The expected CRPS admits the following decomposition,
$$
\tE\{\operatorname{CRPS}(\widehat F)\} = \tE\{D_2^2(\widehat F,F)\} + \tE\{\operatorname{CRPS}(F)\}.
$$
\end{proposition}
Proof is given in the Supplementary Material. In this result, $\tE\{\textrm{CRPS}(F)\}$ can be regarded as an irreducible error, which does not depend on the estimator $\widehat F$. Therefore, minimizing the expected CRPS concerning $\widehat F$ is reduced to minimizing the expected squared conditional $L_2$ discrepancy between $\widehat F$ and $F$. 

As a generic density regression estimator, $\widehat F$ can be constructed by various approaches, including kernel-based methods, spline models, mixture of experts, and Bayesian methods. However, the computational cost is prohibitively high for large samples. To alleviate the computational burden, we consider the density regression problem under a data reduction framework. In detail, we first need to find the best representative data points of size $n$ with minimal empirical squared conditional $L_2$ discrepancy.
Next, we construct an estimator using the representative points as a downstream procedure. Such an estimator ensures that the minimal squared conditional $L_2$ discrepancy is indeed achieved.
Before presenting the algorithm for representative data selection, we introduce the penalized likelihood estimation method in the following subsection.

%%%%%%%%%%%%%%%
%%%%%%%%%%%%%%%

\subsection{Penalized likelihood estimation}\label{sec:ple}

As the downstream modeling strategy, the penalized likelihood method takes the representative data points $\mathbb{D}_n$ of size $n$ as input and estimates a function of interest $\eta$ by minimizing the following criterion 
\begin{equation*}
\ell(\eta \mid \mathbb D_n)+\lambda\, J(\eta),
\end{equation*}
where $\ell(\eta \mid \mathbb D_n)$ is negated log likelihood function of the data as a goodness-of-fit of $\eta$ while functional $J(\eta)$ measures the roughness of $\eta$. The tuning parameter $\lambda$ balances the trade-off between data fidelity and function smoothness.

Conditional density estimation has been studied under the penalized likelihood framework \citep{gu1995smoothing, gu2013smoothing}. The function of interest $f(y \mid \bx)$ is required to be non-negative and integrate to one within its support. A logistic transformation $f(y  \mid  \bx)=e^{\eta(\bx, y)} / \int_{\mathcal{Y}} e^{\eta(\bx, y)}$ is naturally adopted to enforce those constraints, where $\eta(\bx,y)$ is defined on the product domain $\mathcal X\times\mathcal Y$. Moreover, $\eta(\bx, y)$ has to satisfy side conditions to ensure that the transformation is one-to-one, that is, for any $x \in \mathcal{X}$, $A_{y} \eta(\bx, y)=0$, where $A_{y}$ is an averaging operator on $\mathcal{Y}$. An ANOVA decomposition of $\eta$ can be expressed as $\eta(\bx, y)=\eta_{\emptyset}+\eta_{\bx}(\bx)+\eta_{y}(y)+\eta_{\bx, y}(\bx,y)$, and hence $f(y \mid \bx)=e^{\eta_{y}+\eta_{\bx, y}}/\int_{\mathcal{Y}} e^{\eta_{y}+\eta_{\bx, y}}$, where $A_{y}\left(\eta_{y}+\eta_{\bx, y}\right)=0$ for $\bx \in \mathcal{X}$. After eliminating $\eta_{\emptyset}+\eta_{\bx}$ from $\eta(\bx,y)$, we can estimate $f(y  \mid  \bx)$ with $e^{\eta(\bx, y)} / \int_{\mathcal{Y}} e^{\eta(\bx, y)}$ where $\eta(\bx,y)$ minimizes
\begin{equation}\label{smoothingcri}
    -\frac{1}{n} \sum_{i=1}^{n}\left\{\eta\left(\bx_{i}, y_{i}\right)-\log \int_{\mathcal{Y}} e^{\eta\left(\bx_{i}, y\right)}\right\}+\frac{\lambda}{2} J(\eta),
\end{equation}
within the reproducing kernel Hilbert space $\mathcal{H}=\left\{\eta: A_{y} \eta=0, J(\eta)<\infty\right\}$. For \eqref{smoothingcri} to be well defined at $\eta=0$, it is necessary to assume a bounded $\mathcal Y$, which presumably covers all the observed responses. As suggested by \cite{gu1995smoothing}, if the response variable $Y$ has an unknown or unbounded natural support, the estimation in \eqref{smoothingcri} should be interpreted as that of $Y$ restricted in $\mathcal Y$. Noticing that $J$ is indeed a squared semi-norm defined on $\mathcal H$, we have the decomposition as $\mathcal H=\mathcal N_J\oplus \mathcal H_J$. The null space $\mathcal N_J=\left\{g: A_{y} \eta=0, J(\eta)=0\right\}$ is of finite dimension with basis $\left\{\phi_{\nu}\right\}_{\nu=1}^{q}$ while its complement $\mathcal H_J$ is still a reproducing kernel Hilbert space with reproducing kernel $R_J$. The representer theorem \citep{wahba1990spline, gu2013smoothing} does not in general apply to density regression because the likelihood component in \eqref{smoothingcri} relies on an integral term. However, one can still approximate the minimizer in the effective space 
$
\mathcal H^*=\mathcal N_J\oplus \textrm{span}\{R_J((\bx_i, y_i),\cdot): 1\leq i\leq n\}.
$
See \citet{gu1995smoothing} and \citet{gu2013smoothing} for more details of this method, including how to select the tuning parameter.

When the covariate space $\mathcal X$ is multivariate or high-dimensional, it is computationally expensive to repeatedly evaluate the integral $\int e^\eta$. To relieve the burden of numerical integration, \cite{jeon2006effective} devised a penalized pseudo likelihood approach to replace the integral with weighted sums which can be pre-computed. Let $\rho(\bx, y)$ be a known conditional density on the product domain $\mathcal{X}\times \mathcal{Y}$ satisfying $\int_{\mathcal{Y}} \rho(\bx, y)=1$ for $\bx \in \mathcal{X}$. Then the density regression estimator is obtained as $f(y\mid \bx)\propto e^{\eta(\bx,y)}\rho(\bx,y)$ where $\eta(\bx,y)$ minimizes
\begin{equation}\label{smoothingcrifast}
    \frac{1}{n} \sum_{i=1}^{n}\left\{e^{-\eta\left(\bx_{i}, y_{i}\right)}+\int_{\mathcal{Y}} \eta\left(\bx_{i}, y\right) \rho\left(\bx_{i}, y\right)\right\}+\frac{\lambda}{2} J(\eta).
\end{equation}
The penalized pseudo likelihood approach gains great numerical efficiency despite some performance degradation. Algorithms of conditional density estimation under penalized likelihood and pseudo likelihood are available in R package \texttt{gss} \citep{gu2014smoothing}.

Besides its efficient computation, another reason to choose the penalized likelihood estimation as our downstream modeling strategy is the asymptotic properties. In Section~\ref{theoretical}, we show that the expected squared conditional $L_2$ discrepancy between the density regression estimator and the truth is bounded by their symmetrized Kullback-Leibler distance. The convergence rate enjoyed by the penalized likelihood estimator thus implies an error rate of the expected CRPS.

%%%%%%%%%%%%%%%%%%%%%%%%%%%%%%%%%%%%%%%%
\section{Algorithm}\label{algorithm}

\subsection{Conditional support points for density regression}
We now present the procedures to generate conditional support points. Recall that the full data set $\mathbb D_N=\{(\bx_i, y_i)\}_{i=1}^N$ is of size $N$. Our goal is to select conditional support points of size $n$ which represent the true distribution best in terms of the expected CRPS. As pointed in Section~\ref{sec:crps}, Proposition~\ref{prop:crps} implies that such point sets are indeed obtained via minimizing the expected squared conditional $L_2$ discrepancy, $\tE_X\{D_2^2(\widehat F_{n,Y\mid \bX},F_{Y\mid \bX})\}$, where $\widehat F_{n,Y\mid \bX}$ is the empirical conditional distribution function of $n$ data points and $F_{Y\mid \bX}$ is the true conditional distribution function. Furthermore, since $\mathcal Y$ is univariate, it suffices to find point sets minimizing the energy distance in the conditional setting.

In our proposed algorithm, we first partition the covariate space $\mathcal X$ properly, and then conditioning within each partition, we select the points which minimize the conditional energy distance or equivalently the squared conditional $L_2$ discrepancy. Suppose we divide the covariate space $\mathcal X$ into $K$ disjoint partitions, denoted by $B_1,\dots, B_K$. Let $N_k$ be the number of observations with covariates belonging to $B_k$ and $\sum_{k=1}^K N_k=N$. Among the $n$ conditional support points, set $n_k$ to be the number of points allocated in $B_k$ and $\sum_{k=1}^K n_k=n$. 
Extending the relationship between the $L_2$ discrepancy and the energy distance as in \eqref{eq:d2energy} to the conditional distributions, we define the conditional support points $(y^{*(1)}, \dots, y^{*(K)})$ as
\begin{equation}\label{eq:joint}
\argmin_{z^{(1)}, \dots, z^{(K)}}
\biggl\{
\frac{2}{N}\sum_{k=1}^{K} \frac{1}{n_k} \sum_{i=1}^{n_k} \sum_{m=1}^N \left|z_i^{(k)}-y_m\right| 1( \bx_m\in B_k) 
- 
\frac{1}{n}\sum_{k=1}^{K}\frac{1}{n_k} \sum_{i=1}^{n_k}\sum_{j=1}^{n_k}\left|z_{i}^{(k)}-z_{j}^{(k)}\right|
\biggr\},
\end{equation}
where $z^{(k)}=\{z_i^{(k)}\}_{i=1}^{n_k}$ is the $n_k$ points allocated in partition $B_k$.
If $n_k$ is chosen to be proportional to $N_k$, i.e., $n_k=nN_k/N$, we can further simplify the objective function by canceling out a common factor as
$$
\sum_{k=1}^{K} \biggl\{\frac{2}{n_k N_k} \sum_{j=1}^{n_k} \sum_{m=1}^N \left|z_j^{(k)}-y_m\right| 1( \bx_m\in B_k) - \frac{1}{n_k^2} \sum_{i=1}^{n_k}\sum_{j=1}^{n_k}\left|z_{i}^{(k)}-z_{j}^{(k)}\right|\biggr\}.
$$
In this form, quantities within the summation with respect to $k$ are exactly the Monte Carlo approximation of energy distance within each partition. It implies that the joint optimization can be decomposed into $K$ individual sub-problems and leads to parallel computation. In light of this, we present detailed procedures as follows.
\begin{algorithm}\label{alg1}
Conditional support points for density regression.
\begin{enumerate}[Step 1.] 
    \item Covariate space partitioning. Divide the range of each dimension of $\mathcal{X}\in\mathbb{R}^d$ into $\kappa$ disjoint intervals, such that we form a $K=\kappa^d$ partition of $\mathcal{X}$, denoted by $B_1,\dots, B_K$. Then conditioning on $B_k$, the collection of observed data are denoted by $\{(\bx_i^{(k)}, y_i^{(k)})\}_{i=1}^{N_k}$, where $N_k$ is the number of observations with covariates belonging to $B_k$. \label{alg1step0}
    \item Generate support points conditioning on the partitions. In each partition $B_k$, generate conditional support points $y^{*(k)}=\{y_j^{*(k)}\}_{j=1}^{n_k}$ of size $n_k= n N_k/N$. For each point in $y^{*(k)}$, it is then coupled with the corresponding covariates of its nearest neighbor from responses $\{y_i^{(k)}\}_{i=1}^{N_k}$. Denote the coupled data for partition $B_k$ as $(\bx^{*(k)},y^{*(k)})=\{(\bx_j^{*(k)}, y_j^{*(k)})\}_{j=1}^{n_k}$. \label{alg1step1}
    \item Combine the coupled data from all partitions together to form $\mathbb D_n^*=\{(\bx_j^*, y_j^*)\}_{j=1}^n$.
    \item Apply the downstream penalized likelihood estimation \eqref{smoothingcri} or penalized pseudo likelihood estimation \eqref{smoothingcrifast} with $\mathbb D_n^*$ to obtain the density regression estimator. \label{alg1step2}
\end{enumerate}
\end{algorithm}
There is a balance to strike in choosing the number of partitions $K$. When $K$ is small, there are more observations in each cell, and the conditional support points can be better estimated. On the other hand, when coupling each conditional support point with covariate with nearest neighbor method in Step~\ref{alg1step1}, we require a fine partition, equivalently a large $K$, to ensure the observations within a cell have similar covariate values. Step~\ref{alg1step0} applies the binning method to partition the covariate space. One can choose the disjoint intervals on each dimension to be of equal size if no prior knowledge is available. One disadvantage of the binning method is that it suffers from the curse of dimensionality, especially when $d$ is greater than three. In Section~\ref{sec:voronoi}, we further consider several data-driven partitioning strategies.

We recommend in Step~\ref{alg1step1} to choose $n_k$ to be proportional to $N_k$. It is intuitive to allocate more conditional support points to a cell with more observations. More importantly, the joint optimization problem \eqref{eq:joint} can be decomposed into smaller optimizations within individual partitions. Then we can apply the vanilla support points generating method \citep{mak2018support} to each partition in parallel. Numerical comparison with other choices of $n_k$ can be founded in our simulation study.

Note that conditional support points are defined on the real line and are not actually observed. It is thus necessary in Step~\ref{alg1step1} to couple them with covariate for downstream estimation. Specifically, we identify the nearest observed response to each conditional support point and pair its covariate with the latter. It is worth noting that $\mathbb D_n^*$ is synthetic since we use $\{y^{*(k)}\}_{k=1}^K$ as responses rather than the actually observed ones. The reason behind this choice is from both theoretical and optimization perspectives. By definition, conditional support points optimize an energy-distance-based criterion, based on which we establish the optimality of the density regression estimator using $\mathbb D_n^*$ in Section~\ref{theoretical}. Although it is possible to extend this optimality result to a reduced dataset consisting of actual observations, we need to reformulate the optimization problem within the discrete space of all observed responses. However, the state-of-art integer programming techniques are usually slow to find the optimal solution \citep{joseph2021split}. Therefore, it is theoretically provable and computationally efficient to use $\mathbb D_n^*$ for downstream modeling.

For implementation, we suggest using penalized pseudo likelihood estimation in Step~\ref{alg1step2} for fast computation, especially when the dimension of the covariate space is greater than one. 
Regarding the algorithmic complexity, it is known from \cite{mak2018support} that identifying the conditional support points at Step~\ref{alg1step1} takes $O(Nn)$ because the response space $\mathcal Y\subset \mathbb R$. Meanwhile, the coupling procedure involves the nearest neighbor search and requires calculating pairwise distances. However, its cost is negligible since operations are for $\mathbb R$ as well. Once the representative data points of size $n$ are selected, the penalized likelihood estimation in Step~\ref{alg1step2} has computational complexity $O(n^3)$ in general. The representative-point size $n$, as a hyper-parameter, should be chosen based on the available computational resource.

\subsection{Marginal conditional support points for density regression}
When the dimension of $\mathcal X\in\mathbb R^d$ is moderately high, the previous covariate space partitioning will suffer from the curse of dimensionality. When we increase the number of intervals on each dimension, the total number of partitions grows exponentially. Consequently, observations will be scarce in almost all partitions. To address this issue, we propose partitioning only one dimension at a time, obtaining the conditional support points, and repeating on all dimensions of $\mathcal X$. It can be understood as a marginalized version of Algorithm~\ref{alg1}.
\begin{algorithm}\label{alg2}
Marginal conditional support points for density regression. Set $n_{q}\in\mathbb N_0$ as the number of representative points for the $q$th dimension of $\mathcal{X}$, $q=1,\dots,d$ such that $\sum_{q=1}^d n_q=n$. Perform the following steps: 
\begin{enumerate}[Step 1.] 
    \item For each dimension $q$ in $\mathcal{X}$, apply Algorithm~\ref{alg1} to the $q$th dimension of $\mathcal{X}$: first identify the conditional support points $y^{*(q)}=\{y_j^{*(q)}\}_{j=1}^{n_q}$, and then obtain the coupled data $(\bx^{*(q)},y^{*(q)})=\{(\bx_j^{*(q)}, y_j^{*(q)})\}_{j=1}^{n_q}$ for the $q$th dimension. \label{alg2step1}
    \item Combine the coupled data from all dimensions to form $\mathbb D^*_n$. \label{alg2step2}
    \item Apply the downstream method with $\mathbb D^*_n$ to obtain the density regression estimator. \label{alg2step3}
\end{enumerate}
\end{algorithm}
The above marginalization can be carried out in parallel, which helps save computational time greatly. Similar to the nearest neighbor method and the covariate coupling procedure of Algorithm~\ref{alg1}, the above algorithm selects the representative points for each dimension of $\mathcal X$ by first identifying the conditional support points in $\mathcal Y\subset\mathbb R$ restricted in the marginalized direction, and then pairing each conditional support point with covariate whose response value is the closest to the former. The numbers of the representative points for each dimension can be equally allocated, but it is by no means stringent. For instance, when prior knowledge implies that some dimension admits a delicate marginal dependence, one can assign a larger $n_q$ than other dimensions. Algorithm~\ref{alg2} is by construction an approximation to Algorithm~\ref{alg1}, and it is recommended when the dimension of covariate space $\mathcal X$ is equal to or higher than three.

\subsection{Partitioning with a Voronoi tessellation}\label{sec:voronoi}

To deal with the curse of dimensionality when partitioning the multivariate covariate space $\mathcal X\in\mathbb R^d$, we can alternatively apply a Voronoi tessellation. It provides a data-driven partitioning strategy by contrast with the binning method in Step~\ref{alg1step0} of Algorithm~\ref{alg1}. The Voronoi tessellation is defined with $K$ centers $\{\bc_1,\dots,\bc_K\}$ that divide the covariate space $\mathcal X$ into $K$ disjoint regions $B_1,\dots,B_K$, where $B_k$ consists of all the $\bx$'s that are closest to the center $\bc_k$. Formally, if the covariate space $\mathcal X$ is equipped with the Euclidean distance $\|\cdot\|_2$, then $B_k=\{\bx\in\mathcal X: \|\bx-\bc_k\|_2\leq\|\bx-\bc_l\|_2 \textrm{ for all } l\neq k\}$. The Voronoi centers can be selected using clustering based methods such as $k$-means or the vanilla support points \citep{mak2018support}. Once the partition based on a Voronoi tessellation is formed, we proceed with the rest of Algorithm~\ref{alg1} to generate conditional support points for downstream modeling.

We show in Sections~\ref{simulation} and \ref{real} that the aforementioned approaches work well for multivariate covariate space. Under certain scenarios where Algorithm~\ref{alg1} is infeasible, the marginal approach and partitioning with a Voronoi tessellation can both provide reasonable estimate. 

%%%%%%%%%%%%%%%%%%%%%%%%%%%%%%%%%%%%%%%%
\section{Theoretical Properties}\label{theoretical}

\subsection{Distributional convergence}

In this section, we present the distributional convergence of the conditional support points to the desired distribution. Technical proofs of all theoretical results are given in the Supplementary Material. In our proposed procedures, we first partition the covariate space and then choose the points minimizing the energy distance conditioning within each partition. Following this strategy, we first show that the covariate space partitioning yields the marginally distributional convergence.

\begin{lemma}\label{thm:margin}
Let $ \bX \sim F_{ \bX }$ and $ \bX _n\sim F_{n, \bX }$, where $F_{n, \bX }$ stands for the empirical distribution function of the points selected by Algorithm \ref{alg1}. If the number of partitions $K(n)\to\infty$ as $n\to\infty$, then $F_{n,\bX}$ converges almost surely to $F_{\bX}$.
\end{lemma}

This result suggests that the covariate space partitioning indeed provides a good approximation to the true marginal distribution in $\mathcal X$. In order to establish the joint distributional convergence, we need to investigate the convergence of the conditional distribution as well. For $\bx\in \mathcal X$, let $B_{\bx}$ be the partition containing $\bX=\bx$.
Denote by $\phi_{Y \mid  \bX }(s\mid B_{\bx})$ and $\phi_{n,Y \mid \bX}(s\mid B_{\bx})$ the characteristic function of $Y \mid  \bX \in B_{\bx} $ and its corresponding empirical counterpart of conditional support points from Algorithm~\ref{alg1}, respectively. In the following lemma, we prove the convergence of the conditional characteristic function when the
measure of $B_{\bx}$ goes to zero and the number of conditional support points goes to infinity.
Recall that in Section~\ref{sec:ple} we have assumed that $\mathcal Y$ is bounded .

\begin{lemma}\label{thm:cond}
Suppose $\mathcal X$ has a finite measure and $\mathcal Y$ is bounded. The conditional density $f_{Y\mid\bX}$ is uniformly continuous concerning $\bX$ almost surely on $\mathcal{Y}$.
If the number of partitions $K(n)$ and the number of points in each partition goes to infinity as $n\to\infty$, then $\phi_{n,Y \mid \bX}(s\mid B_{\bx})-\phi_{Y \mid  \bX }(s\mid B_{\bx})$ converges to zero almost surely for any $\bx\in\mathcal X$.

\end{lemma}

We now present the main result on the distributional convergence of conditional support points. Let $( \bX ,Y)\sim F$ and $( \bX _n,Y_n)\sim F_n$, where $F_n$ is the empirical distribution function of the conditional support points for density regression via Algorithm \ref{alg1}. 

\begin{theorem}
Suppose $\mathcal X$ has a finite measure and $\mathcal Y$ is bounded. Assume that the marginal density $f_{\bX}$ exists and is bounded away from zero and infinity. Suppose the conditional density $f_{Y\mid\bX}$ is uniformly continuous concerning $\bX$ almost surely on $\mathcal{Y}$. If the number of partitions $K(n)$ and the number of points in each partition goes to infinity as $n\to\infty$, then $( \bX _n,Y_n)$ converges in distribution to $( \bX ,Y)$.
\end{theorem}

This theorem shows that the representative points selected by our method can indeed represent the true joint distribution, as the size of the representative points $n$ increases.

\subsection{Error convergence rate}\label{suberr}

We then investigate the error rate of the density regression estimator with conditional support points in terms of the expected CRPS. By Proposition~\ref{prop:crps}, it suffices to focus on the expected squared conditional $L_2$ discrepancy between the density regression estimator and the truth. 
Our analysis relies on the asymptotic results for the penalized likelihood estimator \citep{gu1995smoothing} where the asymptotic convergence is established with the symmetrized Kullback-Leibler distance. In the main theorem, we show that the expected squared conditional $L_2$ discrepancy can be bounded by the symmetrized Kullback-Leibler distance, and hence follows the convergence rate.

Under the penalized likelihood framework, we express the true conditional density as $f(y  \mid  \bx)=e^{\eta(\bx, y)} / \int_{\mathcal{Y}} e^{\eta(\bx, y)}$, and the estimator as $\widehat f(y  \mid  \bx)=e^{\widehat\eta(\bx, y)} / \int_{\mathcal{Y}} e^{\widehat\eta(\bx, y)}$.
Define $u_\eta(g\mid \bx)=\int_{\mathcal{Y}}g e^{\eta}/\int_{\mathcal{Y}} e^{\eta}$ for $g\in\mathcal H$, and write
$v_\eta(g\mid \bx)=u_{\eta}(g^2\mid \bx)-u_{\eta}(g\mid \bx)^2$.
The symmetrized Kullback-Leibler distance between the density regression estimator and the truth is defined as
$
\text{SKL}(\widehat f_{Y\mid \bX}, f_{Y\mid \bX})=\text{SKL}(\widehat F_{Y\mid \bX}, F_{Y\mid \bX})=\text{SKL}(\widehat \eta, \eta_0)=\tE_{\bX}\{u_{\eta_0}(\eta_0-\widehat\eta\mid \bX)+u_{\widehat\eta}(\widehat\eta-\eta_0\mid \bX) \}.
$
One can show that the quantity inside the expectation operator is exactly the sum of Kullback-Leibler divergences between the estimated and true distributions.

For brevity, we state below the convergence rate results for penalized likelihood estimator \eqref{smoothingcri}, and the similar treatment for penalized pseudo likelihood estimator \eqref{smoothingcrifast} is referred to \cite{gu2013smoothing}. Define $V(\eta-\eta_0) = \int_{\mathcal{X}} v_{\eta_0}(\eta-\eta_0) \,\textrm{d}F_{\bX}(\bx)$.
It can be shown by a first order Taylor expansion that $V(\eta-\eta_0)$ approximates $\text{SKL}(\eta,\eta_0)$ for $\eta$ near $\eta_0$. Regularity conditions are given in the Supplementary Material.

The standard convergence rate of the density regression estimator, e.g., Theorem A.2 of \cite{gu1995smoothing}, is stated as follows, which requires the samples to be independent and identically distributed. Let $\widetilde F_{n,Y\mid \bX}$ denote the estimated distribution function based on independent and identically distributed samples of size $n$ from $F_{Y\mid \bX}$, for any $\bx\in\mathcal X$.
\begin{theorem}\label{gu}
If $\sum_\nu \rho_\nu^p V(\eta_0,\phi_\nu)^2<\infty$ for some $p\in[1,2]$. Under Assumptions 1 to 4 in the Supplementary Material, as $\lambda \rightarrow 0$ and $n \lambda^{2 / r} \rightarrow \infty$, then $\operatorname{SKL}(\widetilde F_{n, Y\mid \bX}, F_{Y\mid \bX})=O_{p}(n^{-1} \lambda^{-1 / r}+\lambda^p)$.
\end{theorem}
We now present our main result on the error rate for the density regression estimator with conditional support points. Unlike those selected by the uniform subsampling, the representative points selected by our method are no longer independent and identically distributed. Our proof is based on the relationship between the $L_2$ discrepancy and the symmetrized Kullback-Leibler distance. Denote the estimated distribution function based on conditional support points for density estimation from Algorithm~\ref{alg1} as $\widehat F_{n,Y\mid \bX}$. Our method selects the representative points by minimizing the squared conditional $L_2$ discrepancy, and our estimator $\widehat F_{n,Y\mid \bX}$ is closer to the truth under the symmetrized Kullback-Leibler distance than $\widetilde F_{n, Y\mid \bX}$, the estimator computed with the uniformly sampled data points. 
\begin{theorem}\label{thm:main}
If $\sum_\nu \rho_\nu^p V(\eta_0,\phi_\nu)^2<\infty$ for some $p\in[1,2]$. Under Assumptions 1 to 4 in the Supplementary Material, as $\lambda \rightarrow 0$ and $n \lambda^{2 / r} \rightarrow \infty$. Then $\tE\{D_2^2(\widehat F_{n, Y\mid \bX}, F_{Y\mid \bX})\}=O_{p}(n^{-1} \lambda^{-1 / r}+\lambda^p)$. In particular, if $\lambda\asymp n^{-r/(pr+1)}$, then the density regression estimator achieves the optimal convergence rate $O_p\{n^{-pr/(pr+1)}\}$.
\end{theorem}
By Proposition~\ref{prop:crps}, this theorem implies that the error convergence rate for the expected CRPS is also $O_p\{n^{-pr/(pr+1)}\}$ despite the irreducible error. 
The parameter $p$ in the theorem quantifies the roughness of the truth $\eta_0$. The roughest $\eta_0$ with $p=1$ satisfies $J(\eta_0)=\sum_\nu \rho_\nu V(\eta_0,\phi_\nu)^2<\infty$, and the optimal convergence rate is $O_p(n^{-r/(r+1)})$. When $\eta_0$ lies in the Sobolev space $\mathcal W_2^m(\mathcal X\times \mathcal Y)$ defined on  a bounded domain in $\mathbb R^{d+1}$, we have $r=2m/(d+1)$, and Theorem~\ref{thm:main} implies the convergence rate is $O_p\{n^{-2m/(2m+d+1)}\}$.

The above result also provides the rationale to use penalized likelihood estimation as the downstream modeling strategy once conditional support points are selected from the full data. Although density regression estimators using the representative points of the same size share the same error convergence rate, our proposal has better empirical performance due to the representative property of conditional support points. 

%%%%%%%%%%%%%%%%%%%%%%%%%%%%%%%%%%%%%%%%
\section{Simulation}\label{simulation}

In the simulation study, we assess the performance of our proposed methods and compare them with other data reduction approaches such as the uniform subsampling and the vanilla support points \citep{mak2018support}. 
Simulated data are generated under various settings. In each simulation, we generate the full data set of size $10^5$, and then randomly divide it into a partition of $95\%$ for training and $5\%$ for testing. Representative data of size $n$ are selected from the training set according to different methods. We set the number of partitions in Algorithms~\ref{alg1} and \ref{alg2} as $K(n)=n^{3/5}$. The response values of observations are normalized. We use the test set to make an out-of-sample evaluation of the CRPS.

We first consider three cases where the covariate space $\mathcal X$ is bivariate.
\begin{enumerate}[\text{Case} 1.]
    \item Covariates $X_1,X_2$ follow $\textrm{Beta}(2,5),\textrm{Beta}(5,2)$. The conditional distribution of $Y$ given $\bX$ is  $\textrm{Beta}(X_1,X_2)$. 
    %2d16
    \item Covariates $X_1$ and $X_2$ are jointly normally distributed with zero means. The covariance is a diagonal matrix of entries $5$. The conditional distribution of $Y$ given $X_1, X_2$ is an exponential distribution with a rate parameter of $(X_1^2+X_2^2)^{1/2}$.
    %2d2
    \item Both covariates follow the uniform distribution on $[0,1]$, the conditional distribution $Y \mid X_1,X_2$ is a mixture of $\textrm{N}(X_1,X_1^2)$ and $\textrm{N}(X_2,X_2^2)$ with mixing probabilities $1-X_1$ and $X_1$, respectively. %2d12
\end{enumerate}
When the conditional density function is transformed with logarithm, the interactions between the response and covariates in Case~1 are separated.
For Cases~2 and 3, there exists only three-way interaction among $Y$, $X_1$ and $X_2$. In particular, in Case~2, the response is isotropic in the covariate space. Case~3 is more complex with response $y$ following the mixture of Gaussian concerning both $X_1$ and $X_2$.
Furthermore, three simulations with trivariate $\mathcal X$ are as follows. They are directly extended from the previous ones.
\begin{enumerate}[\text{Case} 1.]
    \setcounter{enumi}{3}
    \item Covariates $X_1,X_2,X_3$ follow $\textrm{Beta}(2,5), \textrm{Beta}(5,2), \textrm{Beta}(2,2)$, respectively. The conditional distribution $Y \mid \bX$ is $\textrm{Beta}(X_1+X_3,X_2+X_3)$. 
    %3d2
    \item Covariates $X_1$, $X_2$ and $X_3$ are jointly normally distributed with zero means. The covariance is a diagonal matrix of entries $5$. The conditional distribution $Y$ given $\bX$ is an exponential distribution with a rate parameter of $(X_1^2+X_2^2+X_3^2)^{1/2}$.
    %3d7
    \item Each covariate follows the uniform distribution on $[0,1]$, the conditional distribution $Y \mid \bX$ is a mixture of $\textrm{N}(X_1,X_1^2)$, $\textrm{N}(X_2,X_2^2)$ and $\textrm{N}(X_3,X_3^2)$ with mixing probabilities generated from the Dirichlet distribution with parameter $(X_1,X_2,X_3)$.
    %3d6
\end{enumerate}

\begin{figure}[h!]
\centering
\subfigure{
\begin{minipage}[b]{0.3\linewidth}
\includegraphics[width=1\linewidth]{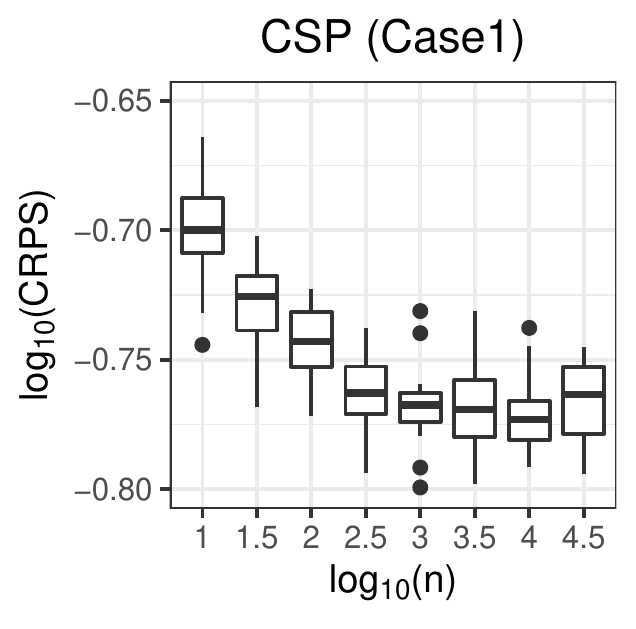}\vspace{4pt}
\includegraphics[width=1\linewidth]{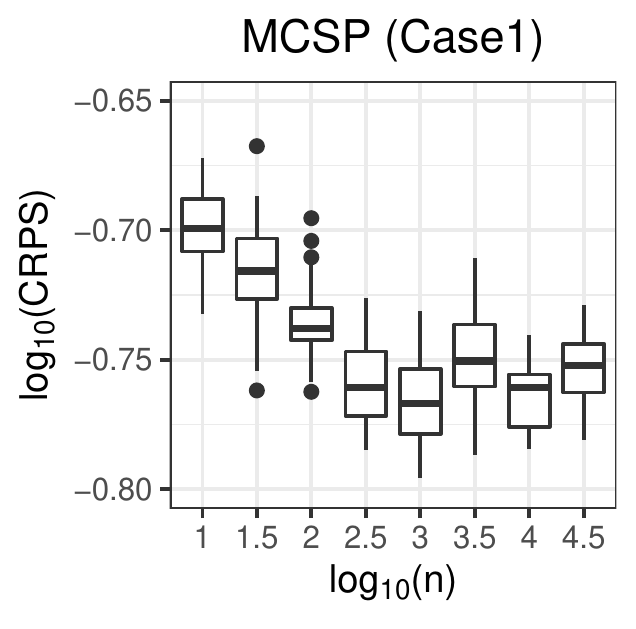}
\end{minipage}}
\subfigure{
\begin{minipage}[b]{0.3\linewidth}
\includegraphics[width=1\linewidth]{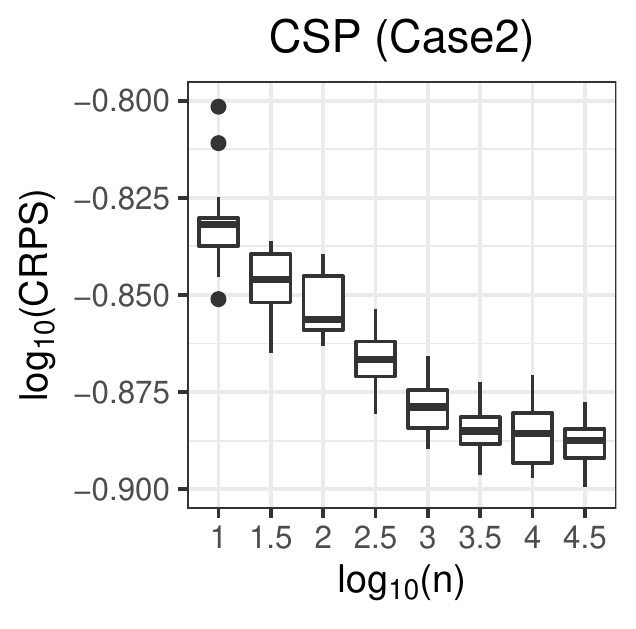}\vspace{4pt}
\includegraphics[width=1\linewidth]{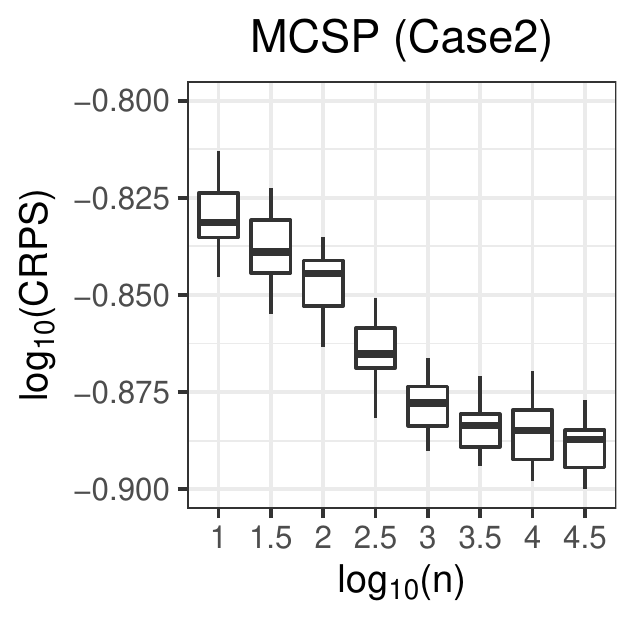}
\end{minipage}}
\subfigure{
\begin{minipage}[b]{0.3\linewidth}
\includegraphics[width=1\linewidth]{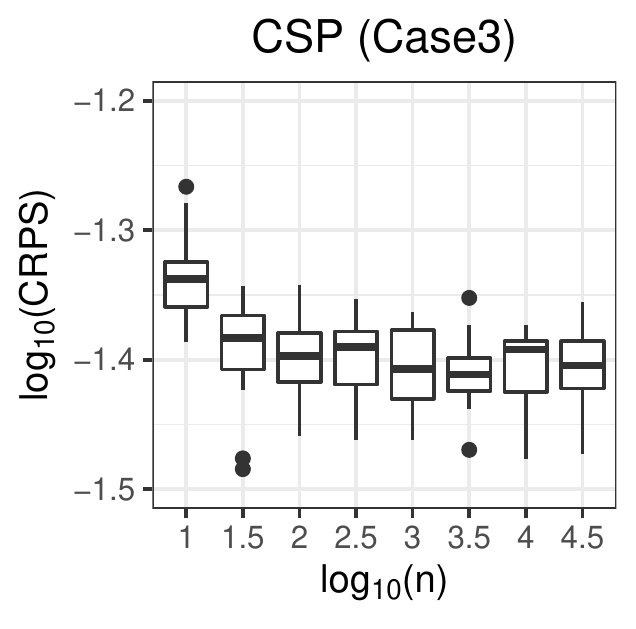}\vspace{4pt}
\includegraphics[width=1\linewidth]{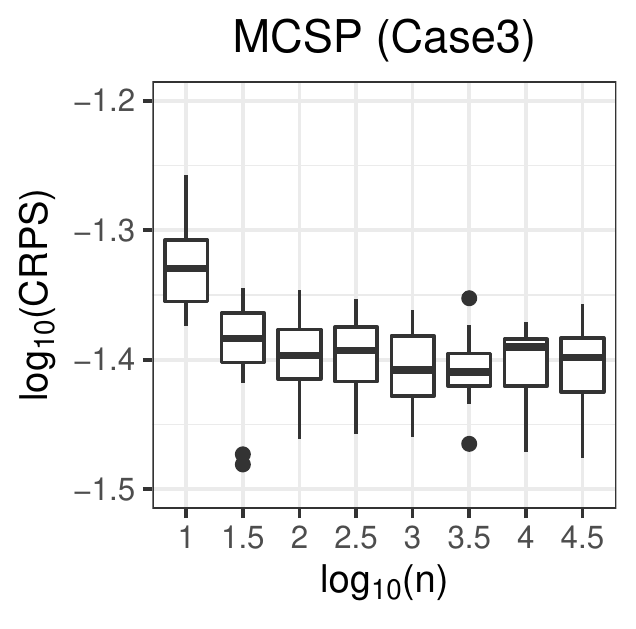}
\end{minipage}}
\caption{Conditional support points (top row) and marginal conditional support points (bottom row) on Cases~1-3: $x$-axis is the logarithm of the representative-point size, $y$-axis is the logarithm of CRPS.}
\label{simurate1}
\end{figure}

\begin{figure}[h!]
\centering
\subfigure{
\begin{minipage}[b]{0.3\linewidth}
\includegraphics[width=1\linewidth]{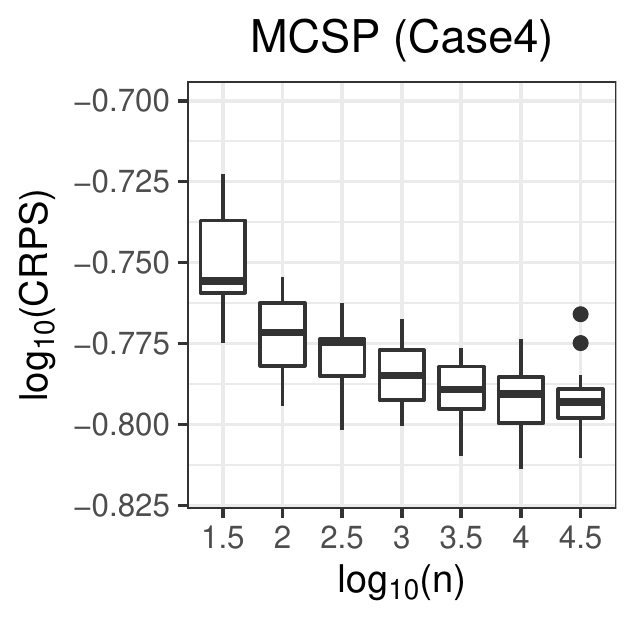}
\end{minipage}}
\subfigure{
\begin{minipage}[b]{0.3\linewidth}
\includegraphics[width=1\linewidth]{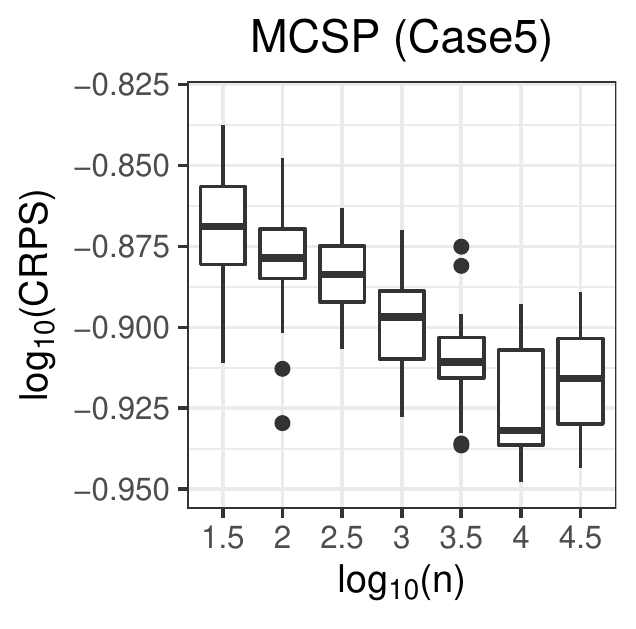}
\end{minipage}}
\subfigure{
\begin{minipage}[b]{0.3\linewidth}
\includegraphics[width=1\linewidth]{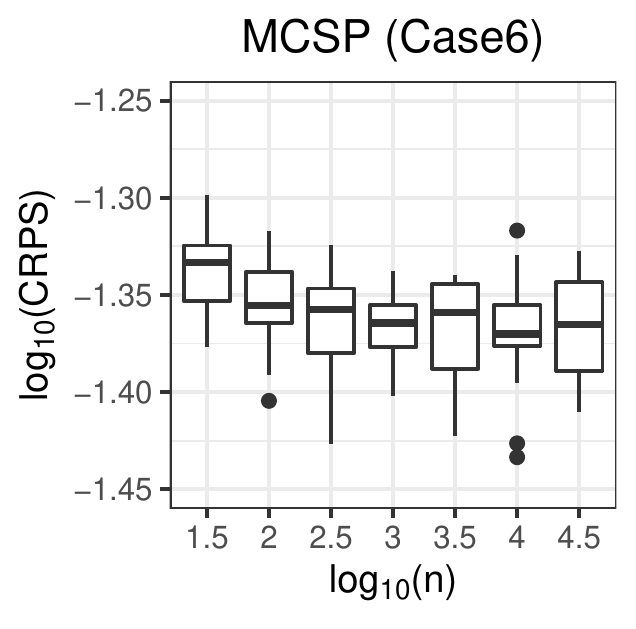}
\end{minipage}}
\caption{Marginal conditional support points on Cases~4-6: $x$-axis is the logarithm of the representative-point size, $y$-axis is the logarithm of CRPS.}
\label{simurate2}
\end{figure}

\subsection{Error convergence rate}

We begin by illustrating the convergence rate of CRPS. Both conditional support points and marginal conditional support points are applied for the bivariate $\mathcal X$ cases, while the latter is implemented for trivariate cases. We repeat each simulation for $20$ times by regenerating the data sets. The size of the representative points ranges from $10$ to $10^{9/2}$ for bivariate cases and $10^{3/2}$ to $10^{9/2}$ for trivariate cases.
Figure~\ref{simurate1} displays the boxplots of CRPS under the logarithmic transformation of different representative-point sizes. First, the logarithm of CRPS starts with a linearly decreasing pattern with the logarithm of the representative-point size, which confirms our error rate analysis in Section~\ref{suberr}. Secondly, as the representative-point size increases, the empirical CRPS converges to a constant level which corresponds to the irreducible error in Proposition~\ref{prop:crps}. Thirdly, according to the comparison between the top and bottom rows in Figure~\ref{simurate1}, the marginal conditional support points indeed provide comparable performance under Cases~1-3, and thus approximates the conditional support points well. It is further applied to trivariate cases. As presented in Figure~\ref{simurate2}, we can observe a similar convergence pattern of CRPS versus the representative-point size.

\subsection{Comparison with other data reduction methods}

Next, we compare our proposed methods with the uniform subsampling and the vanilla support points. 
Both the vanilla support points and our methods are in a deterministic fashion when the full data set and the representative-point size is given. On the other hand, uniform subsampling is a randomized data reduction approach and thus has a positive variance. Therefore, for a fair comparison, we repeat the uniform subsampling for $20$ times after fixing the data set in each simulation. The representative-point size is fixed at $500$ for illustration. A similar conclusion can be drawn by varying the representative-point size.
\begin{figure}[h!]
\centering
\subfigure{
\begin{minipage}[b]{0.3\linewidth}
\includegraphics[width=1\linewidth]{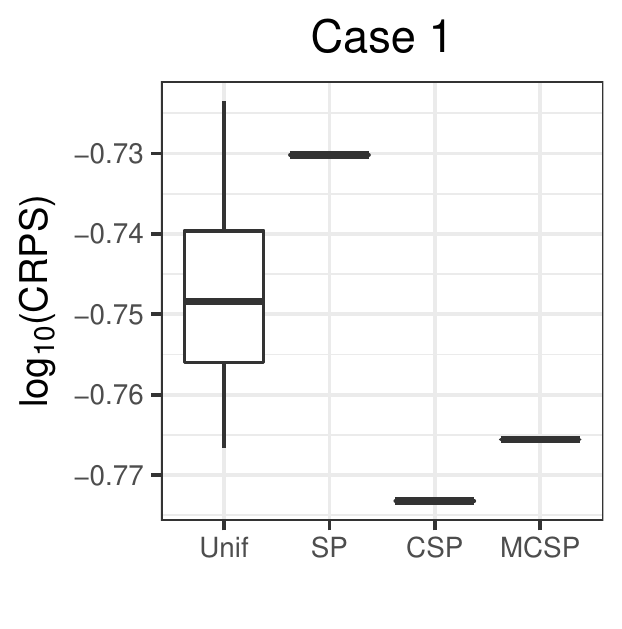}\vspace{4pt}
\includegraphics[width=1\linewidth]{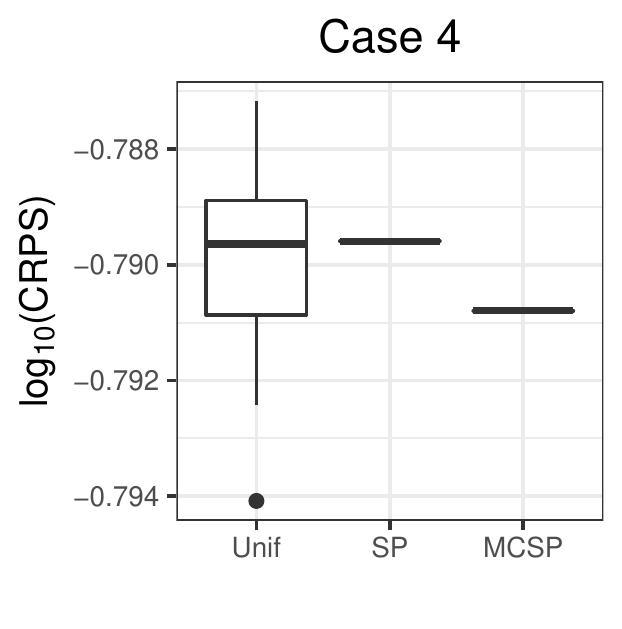}
\end{minipage}}
\subfigure{
\begin{minipage}[b]{0.3\linewidth}
\includegraphics[width=1\linewidth]{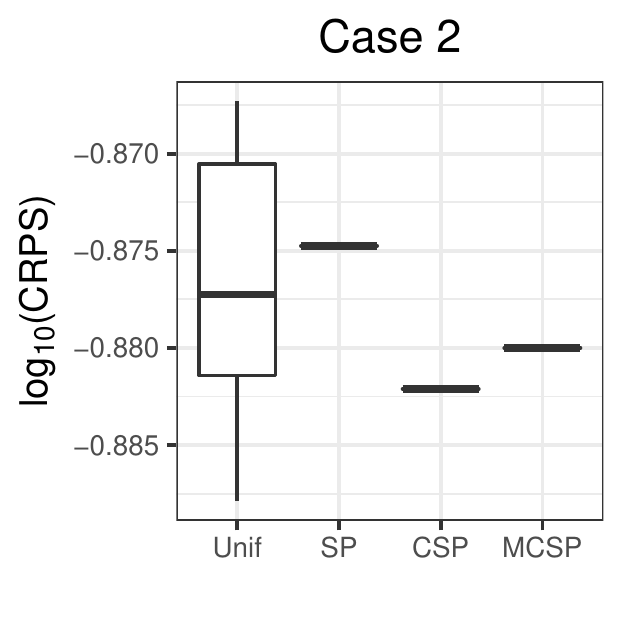}\vspace{4pt}
\includegraphics[width=1\linewidth]{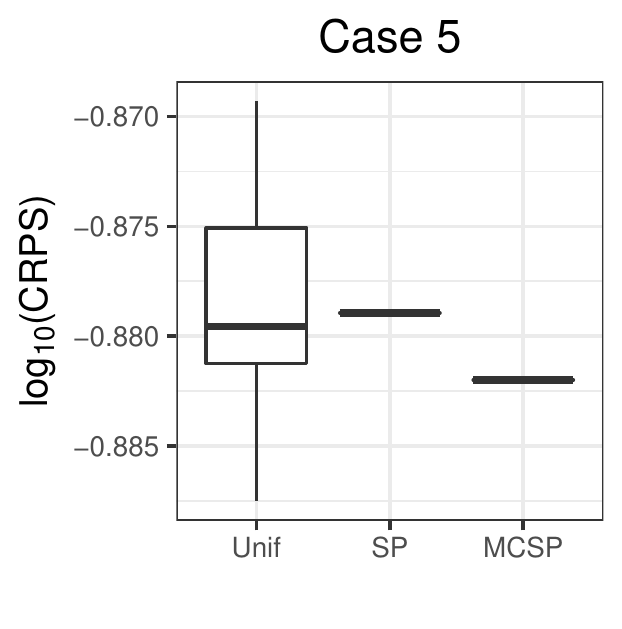}
\end{minipage}}
\subfigure{
\begin{minipage}[b]{0.3\linewidth}
\includegraphics[width=1\linewidth]{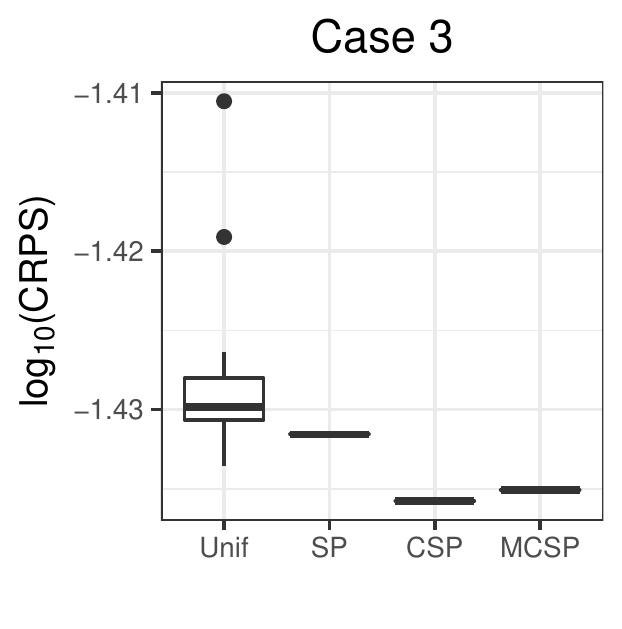}\vspace{4pt}
\includegraphics[width=1\linewidth]{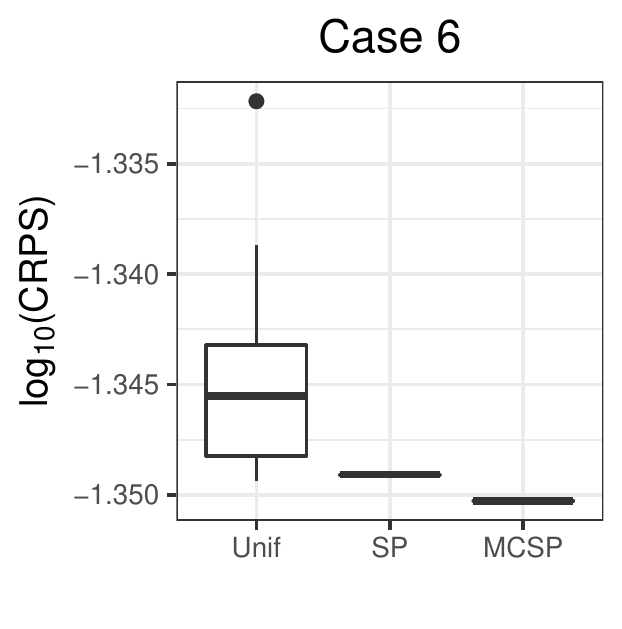}
\end{minipage}}
\caption{Comparison of data reduction methods on Cases~1-6: $y$-axis is the logarithm of CRPS. Unif, SP, CSP and MCSP stand for the uniform subsampling, the vanilla support points and our two proposed methods for density regression.}
\label{simucomp}
\end{figure}
Both conditional support points and its marginal version are applied to Cases~1-3 while only the latter is applied to Cases~4-6. According to Figure~\ref{simucomp}, conditional support points have the best performance among all the methods in bivariate cases, with marginal conditional support points as the runner-up. In the trivariate cases, marginal conditional support points are better than the uniform subsampling and the vanilla support points. Intuitively, the representative points selected by methods based on support points are usually more representative in distribution than the uniform subsampling. The vanilla support points are originally designed for compacting the joint distribution of the response and covariates but are not necessarily the best in the density regression setting.

\subsection{The choice of $n_k$}
\begin{figure}[h!]
\centering
\subfigure{
\begin{minipage}[b]{0.3\linewidth}
\includegraphics[width=1\linewidth]{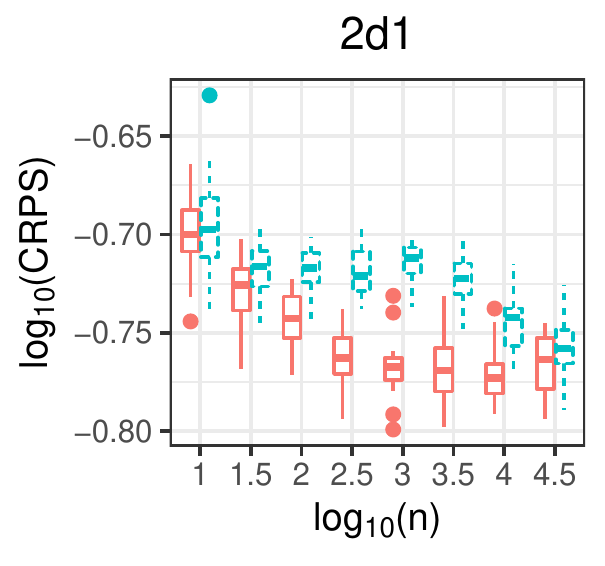}\vspace{4pt}
\includegraphics[width=1\linewidth]{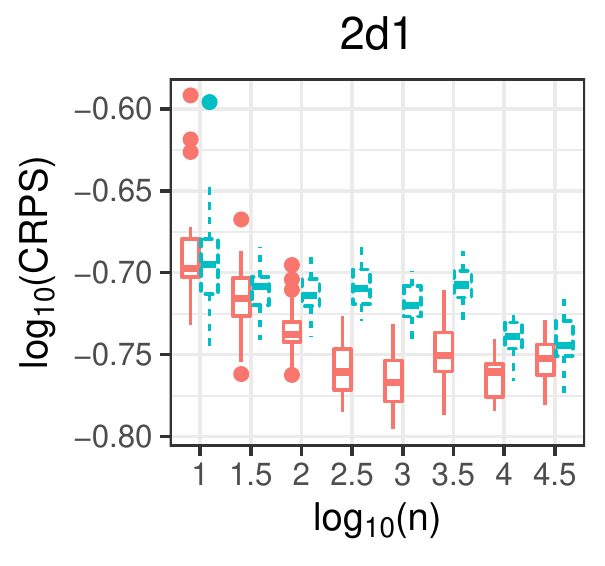}
\end{minipage}}
\subfigure{
\begin{minipage}[b]{0.3\linewidth}
\includegraphics[width=1\linewidth]{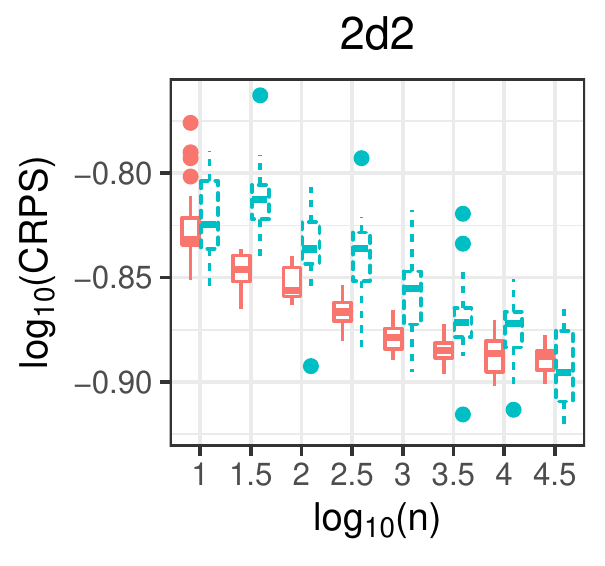}\vspace{4pt}
\includegraphics[width=1\linewidth]{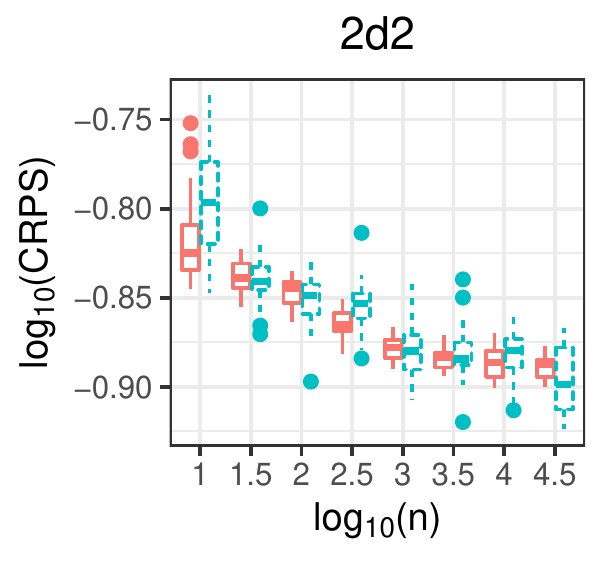}
\end{minipage}}
\subfigure{
\begin{minipage}[b]{0.3\linewidth}
\includegraphics[width=1\linewidth]{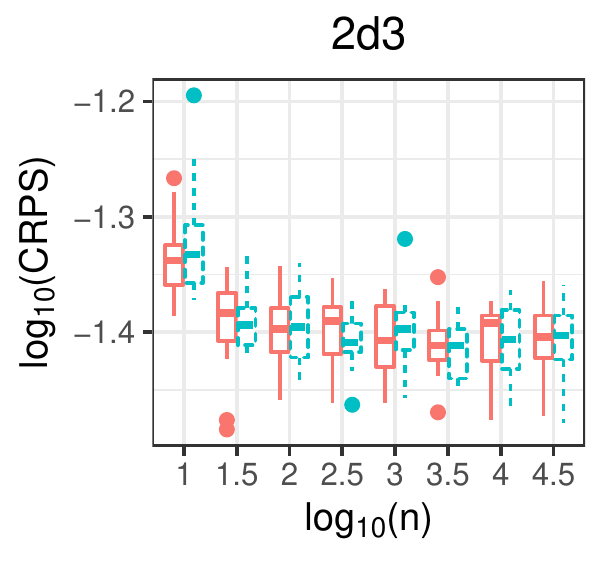}\vspace{4pt}
\includegraphics[width=1\linewidth]{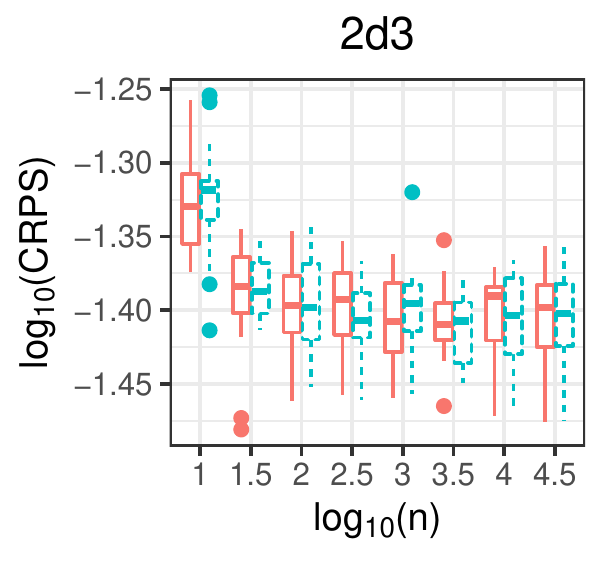}
\end{minipage}}
\caption{Comparison of the CSP (top row) and MCSP (bottom row) with different choices of $n_k$ on three cases of a bivariate covariate space: $n_k$ proportional to $N_k$ (red-solid) and $n_k$ all equal (green-dashed); $x$-axis is the logarithm of the representative-point size, $y$-axis is the logarithm of CRPS.}
\label{simu_compnum}
\end{figure}
We recommend in our conditional support points algorithms that the number of conditional support points in the partition $B_k$ should be proportional to the number of observed data in $B_k$ where $k=1,\dots,K$. It is possible to have other choices of $n_k$, for example, $n_k$'s are all equal across all partitions. We compare both Algorithms~\ref{alg1} and \ref{alg2} with the two choices of $n_k$ on the three cases with a bivariate covariate space. As shown in Figure~\ref{simu_compnum}, the proportional choice $n_k=n N_k/N$ yields a better density regression estimator than the equal choice $n_k=N/K$ in terms of the CRPS, especially in the first two cases. For other simulation examples, we have observed similar results.

\subsection{Covariate space of higher dimension}
\begin{figure}[h!]
\centering
\subfigure{
\begin{minipage}[b]{0.3\linewidth}
\includegraphics[width=1\linewidth]{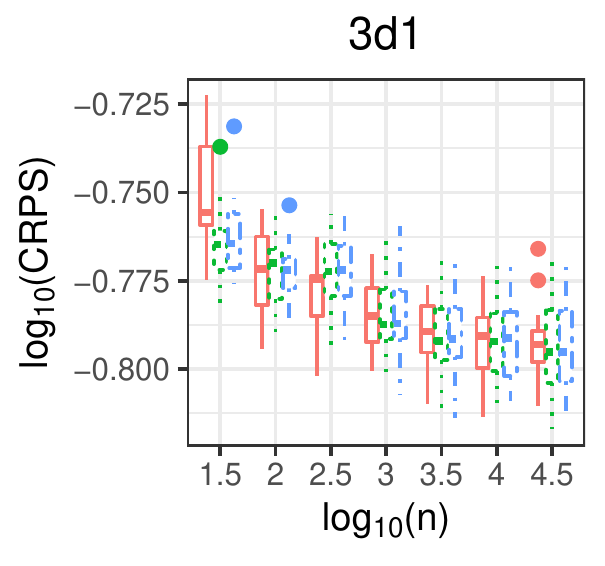}\vspace{4pt}
\includegraphics[width=1\linewidth]{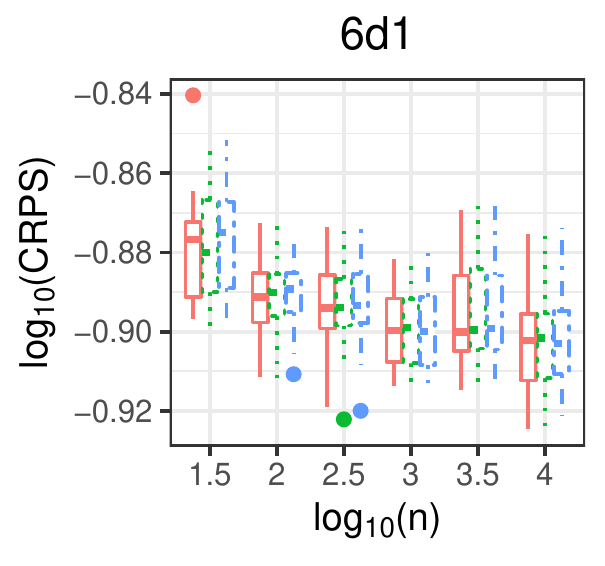}
\end{minipage}}
\subfigure{
\begin{minipage}[b]{0.3\linewidth}
\includegraphics[width=1\linewidth]{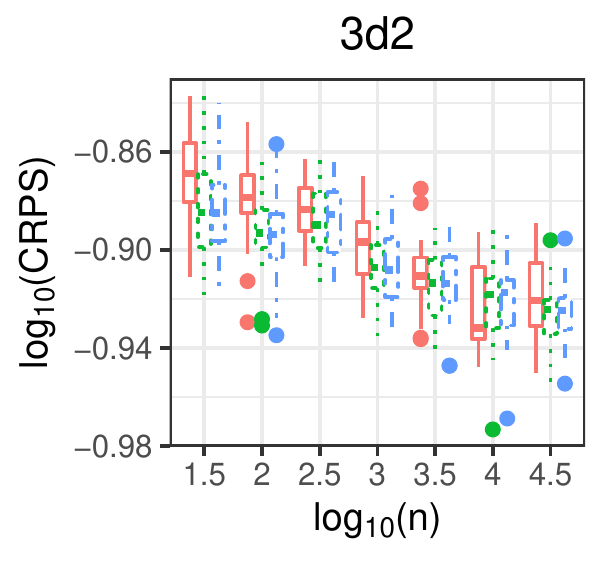}\vspace{4pt}
\includegraphics[width=1\linewidth]{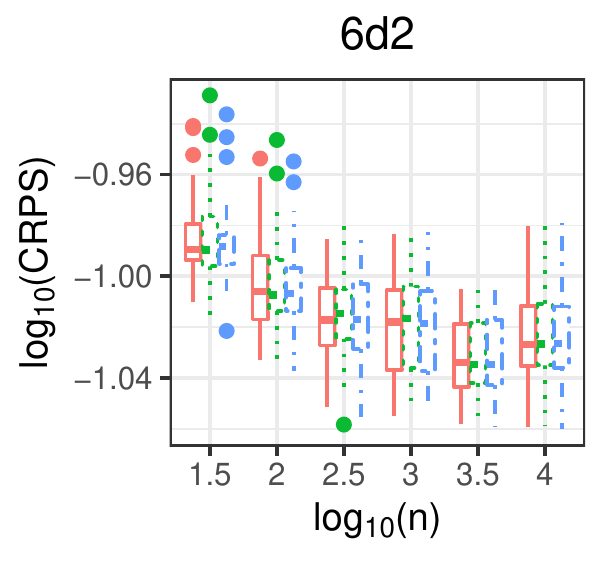}
\end{minipage}}
\subfigure{
\begin{minipage}[b]{0.3\linewidth}
\includegraphics[width=1\linewidth]{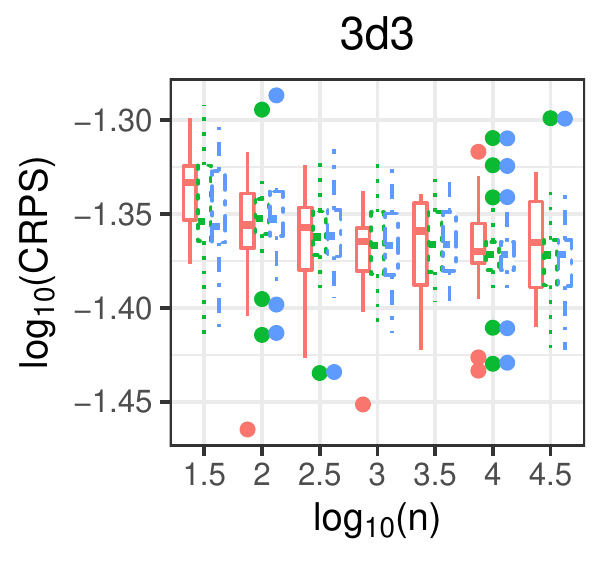}\vspace{4pt}
\includegraphics[width=1\linewidth]{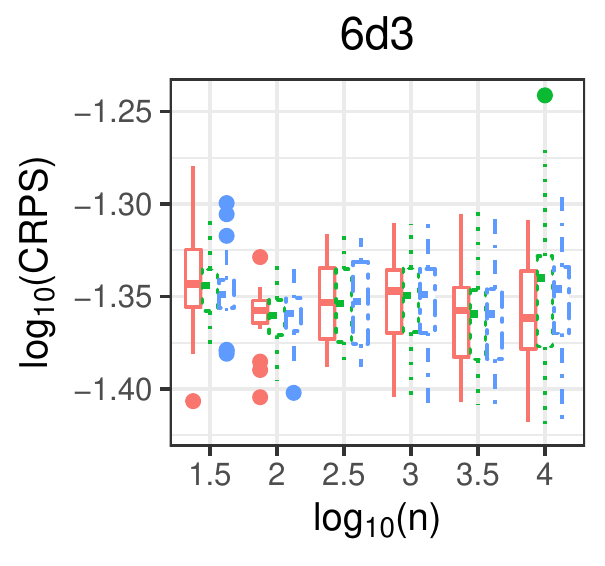}
\end{minipage}}
\caption{Comparison of the MCSP (red-solid) and Voronoi tessellation-based partitioning with $k$-means (green-dotted) and support points (blue-dotdash): $x$-axis is the logarithm of the subsample size, $y$-axis is the logarithm of CRPS.}
\label{simu_compnew}
\end{figure}
In Section~\ref{sec:voronoi}, two partitioning strategies based on a Voronoi tessellation are proposed to mitigate the curse of dimensionality for the binning method in Algorithm~\ref{alg1}. Three additional cases with a six-dimensional covariate space are established by following the same strategy of extending the bivariate covariate space (Cases~1-3) to the trivariate space (Cases~4-6). Their full specifications are omitted for brevity. We compare the marginal conditional support points and two partitioning methods based on a Voronoi tessellation where the Voronoi centers are selected using $k$-means and support points, respectively. Figure~\ref{simu_compnew} shows that the data-driven partitioning methods have comparable performance, and no method dominates the others in terms of the CRPS. According to the decay patterns of the CRPS, it is reasonable to choose the representative-point size to be around $10^{5/2}$ or $10^3$ which yields a satisfactory prediction performance.

\subsection{Computational time}

It is of practical interest to compare the computational time of different upstream data reduction methods when their downstream estimators have similar prediction accuracy. We conducted the experiments on a computer workstation with core Intel Xeon E5-2630 CPU and 62GB RAM. Table~\ref{tab346d} summarizes the averaged computational time of different data reduction methods for cases where the dimension of $\mathcal X$ is three and six. See the Supplementary Material for the comparison for the bivariate covariate space cases. As shown in Table~\ref{tab346d}, the marginal approach is consistently the fastest. Between the two tessellation-based methods, it requires more effort to identify the Voronoi centers by support points than by $k$-means. However, the predictive accuracy of the corresponding downstream estimators does not differ much in Figure~\ref{simu_compnew}. When we choose $n=10^3$ as suggested by Figure~\ref{simu_compnew}, the marginal approach is about five times faster than the other two methods. Note that the Voronoi tessellation is often not feasible for very high dimensions, we recommend to use the marginal approach in cases with a covariate space of moderate dimension.

\begin{table}[h!]
\centering
\footnotesize
\begin{tabular}{crrrrrrr}
\hline\hline
&\multicolumn{3}{c}{$d=3$}	&\multicolumn{3}{c}{$d=6$}	\\
\cmidrule{2-4}\cmidrule{6-8}
$\log_{10}(n)$	&MCSP	&	Voronoi: $k$-means	&	Voronoi: SP&&MCSP	&	Voronoi: $k$-means	&	Voronoi: SP\\
\hline
2	& 329.78 &580.03	&  699.22 	&	&438.83 &  581.94 &  756.25 \\ 
2.5	& 335.62 &1123.20	& 1229.13	&	&445.64 & 1122.17 & 1251.63 \\ 
3	& 470.74 & 2286.96 & 2424.23	&	&470.73 & 2287.55 & 2478.60 \\ 
3.5	& 650.64 & 4427.22 & 4605.84	&	&542.89 & 4426.93 & 4672.49 \\ 
4	& 1008.68 &  9101.91 & 9387.84 &	&1010.15 & 9102.22 & 9513.54 \\ 
4.5	& 2017.00 & 17629.60 & 18096.16 &	&1835.19 &17622.25 &18336.55 \\ 
\hline
\end{tabular}
\caption{Computational time (in seconds) of different upstream data reduction methods for cases with multivariate covariate spaces.}
\label{tab346d}
\end{table}

\section{Application to Wind Turbine Data}\label{real}
Wind energy is one of the fastest-growing sources of electricity generation in the world. According to the wind industry annual market report by the American Wind Energy Association, there are nearly $74$ gigawatts installed wind capacity in the United States, and wind energy provides $4.7\%$ of the total electricity in 2016. In the wind industry, the power curve measures the relationship between turbine power output and the wind speed. It plays a critical role in forecasting wind power \citep{monteiro2009wind, giebel2011state} and assessing turbine performance \citep{albers1999power, stephen2010copula}. 

Besides wind speed, many other environmental factors including wind direction, air density, and wind shear may contribute to changing the distribution of power output. Hence the power curve becomes a power response surface. Technically, modeling the relationship between power output and environmental factors can be understood as a density regression problem. CRPS is used to evaluate the density regression estimator.
In addition to the accuracy criteria, the computational time matters in practice, and any practical solutions need to be reasonably fast. 

In this section, we apply the proposed data reduction method to the large scale wind turbine data set provided by \cite{lee2015power}. There are four inland wind turbines and two offshore turbines. Besides the wind power output $P$, five explanatory variables are available for the four inland wind turbines: wind speed $V$, wind direction $D$, air density $\rho$, turbulence intensity $I$ and below-hub wind shear $S_b$. There are two extra variables for the two offshore wind turbines, namely, humidity $H$ and above-hub wind shear $S_a$. 
Based on previous studies in \cite{jeon2012using} and \cite{lee2015power}, we select the covariates $\bX=(V,D,\rho,I)$ for inland turbines and $\bX=(V,D,\rho,H)$ for offshore turbines, and aim at estimating the conditional density of the response $P$ given $\bX$.

Under the penalized likelihood estimation framework, we estimate $f(y  \mid  \bx)$ via $e^{\eta(\bx, y)} / \int_{\mathcal{Y}} e^{\eta(\bx, y)}$ where the multivariate nonparametric function $\eta(\bx,y)$ can be imposed with ANOVA structure. The physical law of wind power generation \citep{ackermann2005wind,belghazi2012pitch} states that 
$$
P=\frac{1}{2}C_{p}(\beta, \lambda)\rho\pi R^{2}V^{3},
$$
where $R$ is the radius of the rotor, and $C_p$ is called the power coefficient. Although $C_p$ is known to depend on the blade pitch angle $\beta$ and the turbine tip speed ratio $\lambda$, it does not have an analytical formula. Using the partial information from the above physical law and the empirical studies \citep{lee2015power}, we assume the following ANOVA structure for $\eta$ of inland and offshore turbines,
\begin{equation*}
    \begin{aligned}
    \eta_{\text{inland}} & = \eta_{V,P} + \eta_{D,P} + \eta_{\rho,P} + \eta_{I,P} + \eta_{V,D,P},\\
    \eta_{\text{offshore}} & = \eta_{V,P} + \eta_{D,P} + \eta_{\rho,P} + \eta_{H,P} + \eta_{V,D,P}.
    \end{aligned}
\end{equation*}
Since the sizes of the full data for wind turbines are around $10^5$, it is computationally prohibitive to obtain the density regression estimator directly. 
\begin{figure}[h!]
\centering
\subfigure{
\begin{minipage}[b]{0.3\linewidth}
\includegraphics[width=1\linewidth]{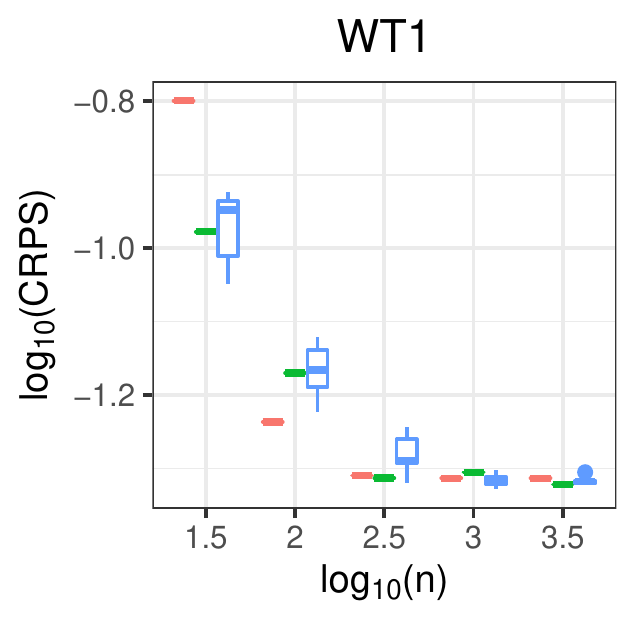}\vspace{4pt}
\includegraphics[width=1\linewidth]{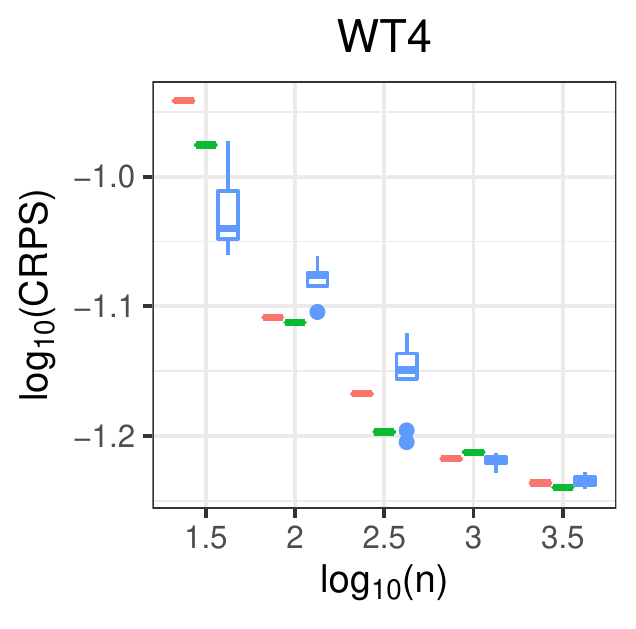}
\end{minipage}}
\subfigure{
\begin{minipage}[b]{0.3\linewidth}
\includegraphics[width=1\linewidth]{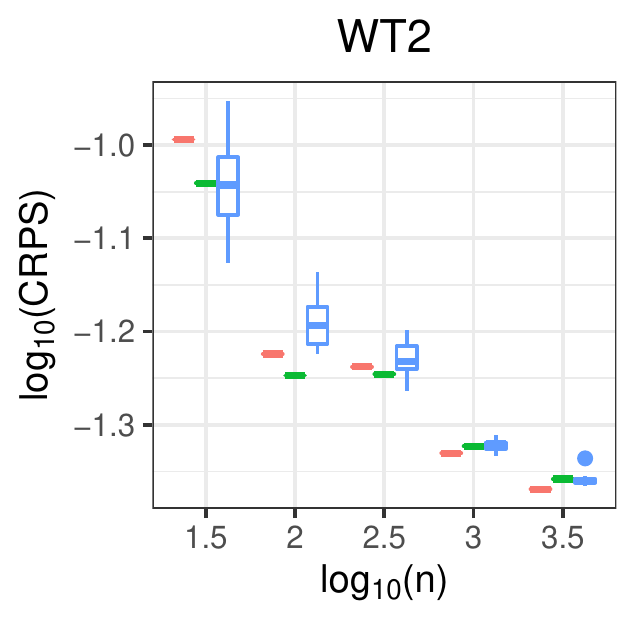}\vspace{4pt}
\includegraphics[width=1\linewidth]{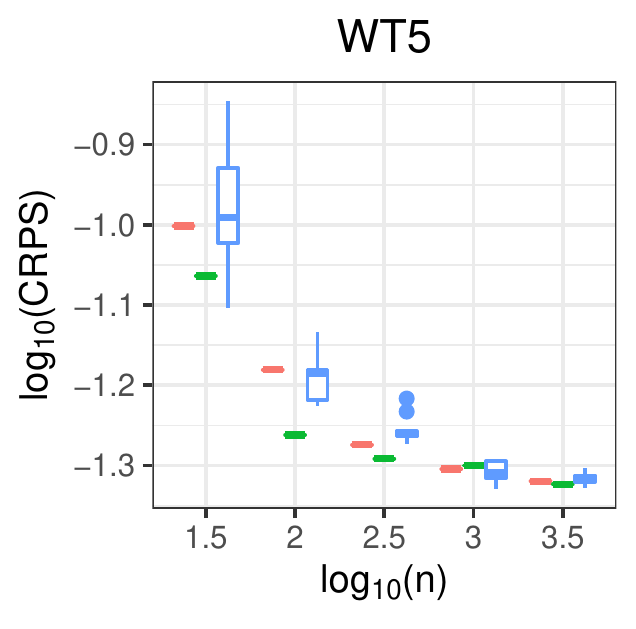}
\end{minipage}}
\subfigure{
\begin{minipage}[b]{0.3\linewidth}
\includegraphics[width=1\linewidth]{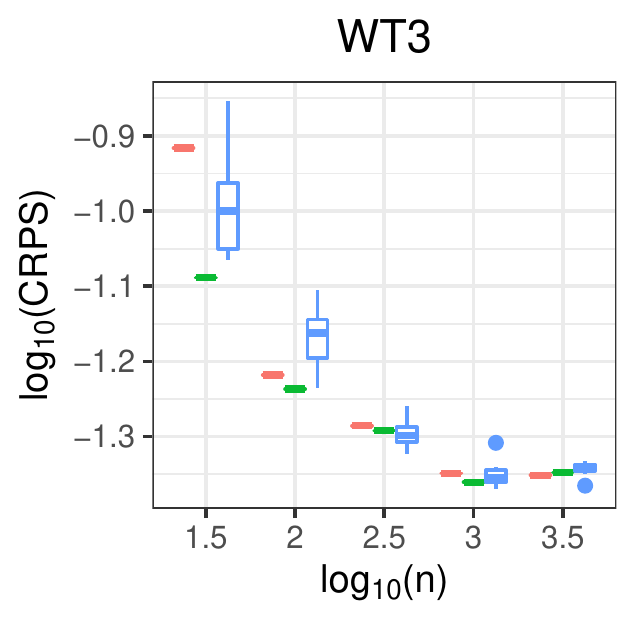}\vspace{4pt}
\includegraphics[width=1\linewidth]{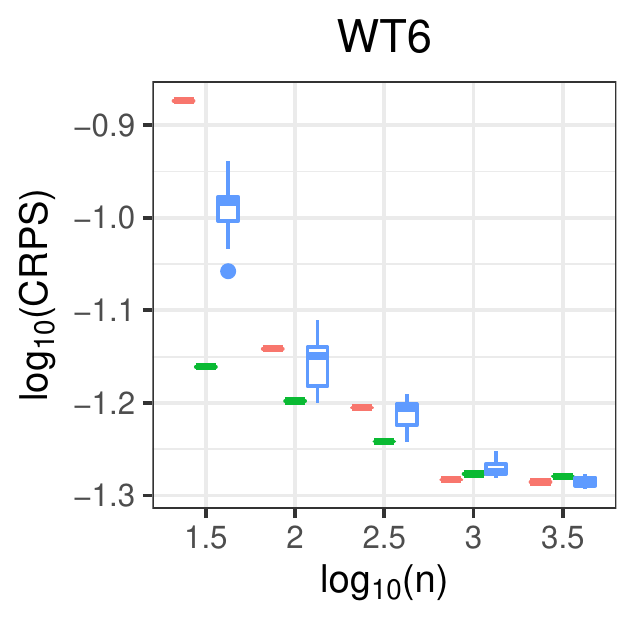}
\end{minipage}}
\caption{Comparison of different methods on wind turbine data (WT1-WT6): $x$-axis is the logarithm of the representative-point size, $y$-axis is the logarithm of CRPS. Within each pair of three boxplots, colored by red, green and blue from left to right, are the vanilla support points, our proposed method, and the uniform subsampling, respectively.}
\label{WT}
\end{figure}

We apply several data reduction methods to the data sets and gather their detailed performance in Figure~\ref{WT}. Different representative-point sizes ranging from $10^{3/2}$ to $10^{7/2}$ are investigated. It is expected that all the methods will perform similarly when the representative-point size is large. Hence, we are more interested in the cases with small or moderate representative-point sizes. For most cases in Figure~\ref{WT}, it is reasonable to choose the representative-point size to be around $10^{5/2}$ or $10^3$ due to a clear pattern of convergence in CRPS. It is clear that our proposed method has the best performance. 

\begin{table}[h!]
\centering
\begin{tabular}{c|cccccc}
\hline\hline
$\log_{10}(n)$&WT1&WT2&WT3&WT4&WT5&WT6\\ \hline
1.5&  287.50&  287.47&  286.69&  286.61&  250.98&  252.50 \\ 
2&  717.72&  718.40&  682.52&  719.12&  574.82&  575.74 \\
2.5& 1368.28& 1368.68& 1331.06& 1430.72& 1149.83& 1149.27 \\
3& 2808.64& 2772.34& 2770.17& 2869.86& 2298.76& 2374.69 \\ 
3.5& 5343.10& 5223.09& 5388.78& 5570.44& 3965.32& 4433.79 \\ \hline
\end{tabular}
\caption{Computational time (in seconds) of the data reduction: wind turbine data.}
\label{tabapp1}
\end{table}

Table~\ref{tabapp1} summarizes the computational time of the upstream data reduction at different representative-point sizes. When the representative-point size is $10^{5/2}$, the data reduction takes about 20 minutes, and the downstream modeling takes less than 10 seconds. As a comparison, directly applying the penalized likelihood estimation to the full data sometimes exhausts the memory storage. Even when an density regression estimator can be computed, it usually takes a few hours. If a timely modeling is desired in practice, we can choose a smaller representative-point size and obtain an estimator in a few minutes, but at the price of prediction accuracy.

%

%%%%%%%%%%%%%%%%%%%%%%%%%%%%%%%%%%%%%%%%
\section{Conclusion}\label{conclusion}

This article developed a novel data reduction approach for density regression with large datasets. The proposed procedure consists of two steps. A set of representative data called conditional support points are first obtained, with which a downstream penalized likelihood density estimation is performed. Using the connection among various distance measures for probability densities, we established the distributional convergence for conditional support points and the optimal rate of convergence for the density regression estimator. Furthermore, efficient algorithms are proposed with numerical experiments illustrating the practical usefulness of the proposed method.

Unlike the original support points \citep{mak2018support}, conditional support points for density regression cannot be fully adapted to the scenario with high-dimensional covariates. This limitation is due to the curse of dimensionality commonly suffered by nonparametric modeling. A possible remedy, as attempted in our analysis of the wind turbine data, is using domain knowledge to guide low-order functional ANOVA structures for the target probability density. It can be an interesting research direction to tailor a partitioning procedure for a specific functional structure.

\bigskip
\begin{center}
{\large\bf SUPPLEMENTARY MATERIALS}
\end{center}

\begin{description}

\item[Supplementary Material] contains technical proofs for theoretical results, and additional numerical results.

\end{description}

\bibliographystyle{Chicago}

\bibliography{bib-power}
\end{document}